\documentclass[
 aip,
 amsmath,amssymb,
 preprint,
]{revtex4-2}

\usepackage[utf8]{inputenc}
\usepackage[dvipsnames]{xcolor}
\usepackage{tikz}
\usepackage{graphicx}
\usepackage{subcaption}
\usepackage{braket}
\usepackage{tabularx}
\usepackage{float}
 
\usepackage{comment}

\usetikzlibrary{shapes.geometric, arrows}
\definecolor{ykred}{rgb}{1.0, 0.0, 0.0}
\definecolor{rygreen}{rgb}{1.0, 0.5, 0.0}
\definecolor{nbblue}{rgb}{0.0, 0.0, 1.0}
\definecolor{csviolet}{rgb}{1.0,0.0,1.0}
\definecolor{jxc}{rgb}{0.2,0.8,0.2}

\begin{document}

\title{Theory of Moment Propagation for Quantum Dynamics in Single-Particle Description}
\author{Nicholas Boyer}
\author{Christopher Shepard}
\author{Ruiyi Zhou}
\author{Jianhang Xu}
\affiliation{Department of Chemistry, University of North Carolina at Chapel Hill, Chapel Hill, North Carolina 27599, USA}
\author{Yosuke Kanai}
\email{ykanai@unc.edu}
\affiliation{Department of Chemistry, University of North Carolina at Chapel Hill, Chapel Hill, North Carolina 27599, USA}
\affiliation{Department of Physics and Astronomy, University of North Carolina at Chapel Hill, Chapel Hill, North Carolina 27599, USA}
\date{\today}

\graphicspath{{Figures/}}
\tikzstyle{startstop} = [rectangle, rounded corners, minimum width=3cm, minimum height=1cm,text centered, draw=black, fill=red!30]
\tikzstyle{io} = [trapezium, trapezium left angle=70, trapezium right angle=110, minimum width=2cm, minimum height=1cm, text centered, draw=black, fill=blue!30]
\tikzstyle{process} = [rectangle, minimum width=10cm, minimum height=1cm, text centered, text width=9cm, draw=black, fill=orange!30]
\tikzstyle{decision} = [diamond, minimum width=3cm, minimum height=1cm, text centered, draw=black, fill=green!30]

\begin{abstract}
We present a novel theoretical formulation for performing quantum dynamics in terms of moments within the single-particle description. By expressing the quantum dynamics in terms of increasing orders of moments, instead of single-particle wave functions as generally done in time-dependent density functional theory, we describe an approach for reducing the high computational cost of simulating the quantum dynamics. 
The equation of motion is given for the moments by deriving analytical expressions for the first-order and second-order time derivatives of the moments, and a numerical scheme is developed for performing quantum dynamics by expanding the moments in the Taylor series as done in classical molecular dynamics simulation. 
We propose a few numerical approaches using this theoretical formalism on a simple one-dimensional model system, for which an analytically exact solution can be derived. Application of the approaches to an anharmonic system is also discussed to illustrate their generality.
We also discuss the use of an artificial neural network model to circumvent the numerical evaluation of the second-order time derivatives of the moments, as analogously done in the context of classical molecular dynamics simulations. 

\end{abstract}

\maketitle

\section{Introduction}

Over the past few decades, explicit time-dependent simulations have become increasingly popular for investigating non-equilibrium quantum dynamics of electrons in chemistry, materials science, and condensed matter physics  \cite{doi:10.1021/acs.chemrev.0c00223, doi:10.1021/acs.jctc.6b00796, doi:10.1021/acs.jctc.9b00729}.  In particular, the real-time propagation approach\cite{PhysRevB.54.4484, Yabana1999, PhysRevB.46.12990} to time-dependent density functional theory\cite{PhysRevLett.52.997} (RT-TDDFT) has significantly advanced over the last decade \cite{doi:10.1063/5.0057587, Kononov2022} for investigating various physical phenomena including optical excitation\cite{ 10.1063/5.0066753, PEMMARAJU201830, Andre_opt}, electronic stopping\cite{PhysRevLett.130.118401,Yi_water,JACS_stopping_per}, charge transfer dynamics\cite{Li_RT_CT,Kummel_CT,doi:10.1021/acsnano.9b08829}, laser-induced water splitting\cite{doi:10.1021/acs.nanolett.1c01187, PhysRevB.99.165424, PhysRevB.96.115451}, electronic circular dichroism \cite{10.1063/5.0038904, 10.1063/1.5128564, 10.1063/1.4953668, doi:10.1021/acs.jctc.9b00632}, and many other electronic excitation phenomena\cite{doi:10.1021/acs.chemrev.0c00223, 10.1063/5.0008194,doi:10.1021/acs.jctc.7b00589}. Within the Kohn-Sham (KS) ansatz, the quantum dynamics of electrons are simulated by propagating a set of single-particle orbitals using time-dependent KS (TD-KS) equations   \cite{doi:10.1063/5.0057587}. The quantum dynamics is invariant with respect to unitary transformation of the single-particle orbitals, and this gauge freedom has been exploited for achieving numerical efficiency of computationally expensive simulations\cite{JIA201921} as well as for gaining physical insights from simulations of complex heterogeneous matter \cite{PhysRevLett.130.118401}. 
 In particular, the gauge transformation into spatially-localized maximally localized Wannier functions (MLWFs)\cite{10.1063/1.5095631, GYGI20031, RevModPhys.84.1419} has proved quite useful in RT-TDDFT simulations not only for efficiently implementing advanced hybrid exchange-correlation approximations \cite{doi:10.1063/5.0057587} but also in studying topological phase in the context of the Floquet engineering\cite{doi:10.1021/acs.jpclett.1c01037, doi:10.1021/acs.jpclett.3c01746}.

In parallel, the emerging field of machine-learning has spurred great interest in employing artificial neural networks (ANNs) for devising efficient schemes for performing quantum dynamics simulations\cite{Quantum_MachineL, doi:10.1021/acs.jctc.2c00702, luo2021autoregressive, Mohseni2022deeplearningof, Rodr_guez_2022, hase2016machine, doi:10.1021/acs.jcim.0c01203, doi:10.1021/acsnano.0c04736}, as ANN models have been effectively utilized in the context of classical molecular dynamics (MD) simulations \cite{DFT_MD, 10.1063/5.0047760, Westermayr_2020, doi:10.1021/acs.jctc.0c01343}. Recent developments have focused on simulating electron dynamics by separating the real and imaginary parts of the wave function \cite{Schiffer_ANN1, Schiffer_ANN2, yao2022emulating, doi:10.1021/acs.jpclett.1c03117}. Secor, et al., for example, introduced a numerical method to propagate wavepackets using ANNs, and proof-of-principle demonstrations were presented in their recent work\cite{doi:10.1021/acs.jpclett.1c03117}. 

These recent developments in the fields motivated us to examine how quantum dynamics based on single-particle wave functions like those of the TD-KS equation can be reformulated as the dynamics of the moments with increasing orders. Some earlier works have considered Heisenberg equation of motion in terms of the position and momentum operators in a related context \cite{10.1063/1.1290288, 10.1063/1.2404677, 10.1063/1.5067005}.
In this work, the equation of motion for moment propagation is derived, and numerical demonstrations are provided for simple proof-of-principle systems, where analytical solutions can be derived. 
We also use an anharmonic potential example to show the generality of the numerical methods based the new theoretical formulation. We further propose how machine-learning approaches like ANN can be conveniently used to perform efficient quantum dynamics simulations with the newly developed moment propagation theory, analogously to the classical molecular dynamics simulation.

\section{Theoretical Formulation}

In this section, we develop the theory of moment propagation. We express single-particle wave functions in terms of moments of increasing orders, and we derive the equation of motion such that Taylor series expansion can be used for propagating the moments in time. 
This formalism allows us to devise a numerically convenient method for performing quantum dynamics simulation using the single-particle description like the TD-KS equation in RT-TDDFT.

\subsection{Equation of Motion for Moments}

We consider the situation in which  quantum dynamics of a many-particle system is described by the propagation of single-particle wave functions via Schrodinger-like equation, as done through formulating the TD-KS equation. 
In order to keep the discussion centered on the theoretical formulation, lengthy mathematical derivations are given in Appendix.
Instead of expressing a single-particle wave function as $\psi(\mathbf{r}, t)$, we explicitly write it as $\psi(x, y, z, t)$, since the orders of moments are generally not the same in the three Cartesian coordinates. The central quantity for our theoretical formulation is the moments of single-particle probability density, given by

\begin{equation}
\braket{x^ay^bz^c}(t)=  \int \int \int {x^ay^bz^c n(x, y, z,t)d x d y d z},
\end{equation}
where $a,b,c$ are non-negative integers used to denote the a-th, b-th, c-th moment in $x,y,z$ directions of the Cartesian coordinate space and $n(x, y, z, t)$ is the particle (probability) density, which is the square modulus of the single-particle wave function (i.e. $n(x, y, z,t)=| \psi(x, y, z, t)|^2$). Moments are widely used as a statistical quantity to characterize features of probability functions. 
In order to propagate moments, the time derivatives of the moments need to be derived from time-dependent Schrodinger equation (TD-SE) or alike,

\begin{equation}
\label{eq:tdse}
i\frac{\partial \psi(x, y, z, t)}{\partial t}  = -\frac{1}{2}\nabla^2 \psi(x, y, z, t) + V(x, y, z, t)\psi(x, y, z, t)
\end{equation}
where $\psi$ is the time-dependent wave function and $V$ specifies time-dependent potential. The first-order time derivative of the moments is given by (see Appendix \ref{sec:eq3derivation})

\begin{equation}
\begin{aligned}
\label{eq:firstder}
\frac{d \braket{x^ay^bz^c}(t)}{d t} =&\int \int \int {x^ay^bz^c \dot n(x, y, z, t)d x d y d z}\\
= & -\frac{i}{2} \int \left[\nabla^2\left(x^ay^bz^c\right)n + 2\nabla\left(x^ay^bz^c\right)\cdot\nabla\psi\psi^*\right]d^3r\\
= & -i \int \int \int (a x^{a-1}y^bz^c \partial_x\psi\psi^*+ b y^{b-1}x^az^c \partial_y\psi\psi^* + c z^{c-1}x^ay^b \partial_z\psi\psi^* \\
+ &\frac{a(a-1)}{2} x^{a-2}y^bz^c n(x, y, z, t) + \frac{b(b-1)}{2} y^{b-2}x^az^c n(x, y, z, t) \\ 
+ &\frac{c(c-1)}{2} z^{c-2}x^ay^b n(x, y, z, t) 
) dx dy dz.
\end{aligned}
\end{equation}
Here, we used the simplified notation such that
$ \partial_g \equiv \frac{\partial}{\partial g}$
for brevity. Similarly, the second-order time derivative of the moments (see Appendix \ref{sec:eq4derivation}) is 
\begin{equation}
\begin{aligned}
\label{eq:secondder}
\frac{d^2 \braket{x^ay^bz^c}(t)}{d t^2} = & \int \operatorname{Re} \left[ - \nabla (x^ay^bz^c) \cdot \nabla V n  + \frac{1}{4} \nabla^4\left(x^ay^bz^c\right)n  -  ( \nabla \otimes \nabla \left(x^ay^bz^c\right)  \cdot (\nabla \otimes \nabla \psi))  \psi^* \right] d^3r\\
= & \operatorname{Re} (
- \int \int \int (ax^{a-1}y^bz^c \partial_x V(x, y, z, t) n(x, y, z, t) \\
+ & b x^ay^{b-1}z^c \partial_y V(x, y, z, t) n(x, y, z, t) + cx^ay^bz^{c-1} \partial_z V(x, y, z, t) n(x, y, z, t))dxdydz\\ 
- & \int \int \int (a(a-1) x^{a-2}y^bz^c  \partial_x^2 \psi  \psi^* + 2ba x^{a-1}y^{b-1}z^c  \partial_x \partial_y \psi \psi^* \\
+ & 2ca x^{a-1}y^bz^{c-1} \partial_x \partial_z \psi  \psi^* + b(b-1) y^{b-2}x^az^c \partial_y^2 \psi \psi^*\\
+ & 2cb y^{b-1}x^az^{c-1}  \partial_y \partial_z \psi  \psi^* + c(c-1) z^{c-2}x^ay^b \partial_z^2 \psi \psi^*
)dx dy dz\\
+ & a(a-1) (\frac{(a-2)(a-3)}{4} \braket{x^{a-4}y^bz^c} + \frac{b(b-1)}{2} \braket{y^{b-2}x^{a-2}z^c} \\ 
+ &\frac{c(c-1)}{2} \braket{z^{c-2}x^{a-2}y^b} ) 
+ b(b-1) (\frac{(b-2)(b-3)}{4} \braket{y^{b-4}x^az^c}\\
+ &\frac{c(c-1)}{2} \braket{z^{c-2}x^ay^{b-2}} ) 
+ \frac{c(c-1)(c-2)(c-3)}{4} \braket{z^{c-4}x^ay^b} )\\
\end{aligned}
\end{equation}
where the shorthand notation $ \partial_{g} \partial_{r} \equiv \frac{\partial^2}{\partial g \partial r}$ is used for brevity. 
Higher order derivatives can be obtained in principle, but their utility in practice is limited because of the mathematical complexity of the resulting expressions. In order to remove explicit reference to the single-particle wave function in Eqs. \ref{eq:firstder} and \ref{eq:secondder}, we still need to express the wave function and its derivatives in terms of the moments. 

\subsection{Single-particle Wave Function and Moments } 
\label{sub:spwf-moment}
The expressions for the first-order and second-order time derivatives of the moments (Eqs. \ref{eq:firstder} and \ref{eq:secondder}) contain explicit dependence on the wave function. We show here that the explicit dependence can be removed by expressing the wave function in the polar form 
\begin{equation}
\label{eq:polar}
\psi(x, y, z,t) = \sqrt{n(x, y, z, t)} e^{i\theta(x, y, z,t)}
\end{equation}
where $n(x, y, z, t)$ is the particle density and $\theta(x, y, z, t)$ represents the phase. 

\subsubsection{Particle density using cumulants via Edgeworth series}
\label{subsubsec:density}
Instead of using the moments directly, it is convenient to transform them into the cumulant representation so that the Edgeworth series \cite{BookonEdgeworth} can be employed for easily evaluating properties such as the particle density.  While the following formalism is applicable in three-dimensions, we focus here on one-dimensional case, $x$, for clarity.
The cumulants are given in terms of the moments as
\cite{10.2307/2684642}
\begin{equation}
\kappa_a=\braket{x^a}-\sum_{i=1}^{a-1}{a-1 \choose i-1}\kappa_i \braket{x^{a-i}}
\end{equation}
where $\kappa_a$ is the a-th cumulant and $\braket{x^a}$ is the a-th moment.
We express the particle density using cumulants via the Edgeworth series by expanding in the basis of Cartesian Gaussian functions\cite{Blinnikov_1998, 10.2307/2237255}. We do so by first generating the n-th order probability density function 
\begin{equation}
\label{eq:pdfn}
pdf_n(x) = z(x)\Big(1+ \sum^{n-2}_{j=1} (\sum^{\{P_j\}}_{\{P_{j_m}\}} \big(\prod_{i=1}^{\infty} \frac{1}{k_i !} (\frac{\lambda_{i+2}}{(i+2 )!})^{k_i}\big) He_s(x))\Big)
\end{equation}
where $\{P_j\}$ is the set of all positive integer partitions of $j$. Integer partitions of an integer $j$ are all possible ways to add positive integers to $j$. For example, if $j=3$, then the integer partitions include $1+1+1=3$ and $1+2=3$. 
For each $P_{j}$, these combinations are denoted by $P_{j_m}$ where m is the index of the partition in the set. For each $m$, Eq. \ref{eq:pdfn} contains several terms.
First, $\lambda_n$ is given by
\begin{equation}
\label{eq:lambda}
\lambda_n = \frac{\kappa_n}{(\sqrt{\kappa_2})^n}
\end{equation}
where $\kappa_n$ is the n-th order cumulant.
Secondly, integer $k_i$ in Eq. \ref{eq:pdfn} is the count of $i$ within each integer partition such that $j = \sum_i i k_i$.
For instance, for the case of the integer partition $P_{j=3_{m=1}}$:$1+1+1$, we have $k_1=3$, $k_{\ge 2}=0$.
For the integer partition $P_{j=3_{m=2}}$:$1+2$, $k_1=1$, $k_2=1$, and $k_{\ge 3}=0$.
Based on the integer value $k_i$, we obtain another integer value $s$ as
\begin{equation}
s = j + 2 \sum_{i=1}^{\infty} k_i.\\
\end{equation}
For example, for the integer partition, $P_{j=3_{m=1}}$:$1+1+1 (i=1)$, we have $s=9$.
The Hermite polynomial $He_s(x)$ in Eq. \ref{eq:pdfn} is then given by
\begin{equation}
He_s(x) = s! \sum^{\lfloor \frac{s}{2} \rfloor}_{j=0}\frac{(-1)^jx^{n-2j}}{2^j(n-2j)! j!}
\end{equation}
where $\lfloor s/2 \rfloor$ denotes the floor function of $s/2$\cite{https://doi.org/10.48550/arxiv.1901.01648}. We use Gaussian functions, $z(x) = \frac{1}{\sqrt{2 \pi}} \exp{(-\frac{x^2}{2})}$, as
the basis in Eq. \ref{eq:pdfn}. The Edgeworth expansion then gives the particle density as

\begin{equation}
\label{eq:pdf}
n(x) = \frac{pdf(\frac{x - \kappa_1}{\sqrt{\kappa_2}})}{\sqrt{\kappa_2}}
\end{equation}
where the probability density function $pdf(x)$ is given by a n-th order probability density function $pdf_n(x)$ using Eq. \ref{eq:pdfn}. 
This procedure effectively expresses the particle density, $n(x)$, from the moments $\braket{x^a}$. 

\subsubsection{Phase from the moments}
\label{subsubsec:phase}
In addition to the particle density $n(x,y,z,t)$ as discussed above, the phase $\theta(x,y,z,t)$ (which is uniquely determined up to a constant) in Eq. \ref{eq:polar} or rather its spatial derivatives need to be related to the moments\cite{PhysRevLett.107.230403}. 
The derivative of the phase with respect to a particular coordinate $u$  can be written (see Appendix \ref{sec:eq13derivation}) as 

\begin{equation}
\label{eq:wfphase}
\frac{\partial \theta(x, y, z,t) }{\partial u} = 
-n(x, y, z,t)^{-1}
\int_{-\infty}^{u }\alpha_{u}(\mathbf{r'},t) du' 
\end{equation}
where $\alpha_u \equiv \frac{i}{2} (\frac{\partial^2 \psi}{\partial u^2}\psi^* - \frac{\partial^2 \psi^*}{\partial u^2}\psi) $ is used. Here, the integration variable $u'$ corresponds to the same coordinate as $u$, and we used $\textbf{r}'$ to indicates that the function depends on all three coordinates $\{x, y, z\}$. 
The phase can be obtained as 
\begin{equation}
\label{eq:theta}
\theta(x, y, z, t) = \phi + \int_0^x \frac{\partial \theta(u, 0, 0,t) }{\partial x} du + \int_0^y \frac{\partial \theta(x, u, 0,t) }{\partial y} du + \int_0^z \frac{\partial \theta(x, y, u,t) }{\partial z}du
\end{equation}
where $\phi$ is an arbitrary constant. 
It is also useful to note that the time derivative of the particle density can be expressed in terms of these functions as $\dot n = \alpha_x + \alpha_y + \alpha_z$ (see Appendix \ref{sec:eq15derivation}).
Instead of the wave function, $\alpha_u$ can be written alternatively using the moment time derivatives (see Appendix \ref{sec:eq15derivation}) as 

\begin{equation}
\label{eq:alphax}
    \alpha_u (x,y,z,t)=  \sum_{a,b,c} c_u^{abc}(t) \frac{d \braket{x^ay^bz^c}}{d t}
    \frac{\partial n}{\partial \braket{x^ay^bz^c}}
\end{equation}
where $c_u^{abc}(t)$ are the coefficients for the moment  $\braket{x^ay^bz^c}(t)$, and these coefficients satisfy $c_x^{abc}(t)+ c_y^{abc}(t)+ c_z^{abc}(t)=1$ by construction.
The product of $c_u^{abc}(t)$ and the time derivative of the moments in Eq. \ref{eq:alphax} are propagated together. For example, one has (see Appendix \ref{sec:eq16derivation})

\begin{multline}
\label{eq:secondderMPT2upcx}
\frac{d}{d t}(c_x^{abc}(t) \frac{d \braket{x^ay^bz^c}(t)}{d t}) =
- \int \int \int (ax^{a-1}y^bz^c\partial_x V(x, y, z, t) n(x, y, z, t))dxdydz \\
+\int \int \int (
a(a-1) x^{a-2}y^bz^c  (\frac{(\partial_x n(x, y, z, t)^2)}{4n(x, y, z, t)} + \frac{L_x(x, y,z,t)^2}{n(x, y, z, t)})
 \\+ ba x^{a-1}y^{b-1}z^c  (\frac{(\partial_x n(x, y, z, t)\partial_y n(x, y, z, t))}{4n(x, y, z, t)} + \frac{L_x(x, y,z,t)L_y(x, y,z,t)}{n(x, y, z, t)}) \\
 +ca x^{a-1}y^bz^{c-1}  (\frac{(\partial_x n(x, y, z, t)\partial_z n(x, y, z, t))}{4n(x, y, z, t)} + \frac{L_x(x, y,z,t)L_z(x, y,z,t)}{n(x, y, z, t)})) dxdydz\\
- a(a-1) (
\frac{(a-2)(a-3)}{4} \braket{x^{a-4}y^bz^c} 
+ \frac{b(b-1)}{4} \braket{y^{b-2}x^{a-2}z^c}+ \frac{c(c-1)}{4} \braket{z^{c-2}x^{a-2}y^b} ) \\
\end{multline}
where we define $L_{x}(x, y, z, t)$ as 
\begin{equation}
\label{eq:Lrup}
\begin{aligned}
L_{x}(x, y, z, t) \equiv - \int_{- \infty}^{x}\alpha_{x}(\mathbf{r'}, t) dx'.\\
\end{aligned}
\end{equation}

\noindent
Note that the time derivative of the coefficients $c_u^{abc}(t)$ themselves are not needed as long as the product of the coefficient and the time derivative of moments are propagated together as can be done using Eq. \ref{eq:secondderMPT2upcx} since they enter together in Eq. \ref{eq:alphax} for obtaining $\alpha_u(t)$.

The key point of this section is 
that the single-particle wave function, expressed as in Eq. \ref{eq:polar}, can be given in terms of the moments and their first-order time derivative. 

\subsection{Time Derivatives of Moments} 
Combining the results of the two preceding Subsections \ref{subsubsec:density} and \ref{subsubsec:phase}, we can now write the first-order and the second-order time derivatives of the moments (Eqs. \ref{eq:firstder} and \ref{eq:secondder}) without referring to the single-particle wave function.
The first-order time derivative of the moments is 
\begin{equation}
\label{eq:1stM}
\frac{d \braket{x^ay^bz^c}(t)}{d t} = \int \int \int {x^ay^bz^c (\alpha_x(x, y, z, t) + \alpha_y(x, y, z, t) + \alpha_z(x, y, z, t) )d x d y d z}.
\end{equation}
\noindent
The second-order time derivative of the moments is (see Appendix \ref{subsec:eq19})
\begin{multline}
\label{eq:secondderMPT2up}
\frac{d^2 \braket{x^ay^bz^c}(t)}{d t^2} =
- \int \int \int (ax^{a-1}y^bz^c\partial_x V(x, y, z, t) n(x, y, z, t) \\
+ b x^ay^{b-1}z^c\partial_y V(x, y, z, t) n(x, y, z, t)
+ cx^ay^bz^{c-1}\partial_z V(x, y, z, t) n(x, y, z, t))dxdydz\\ 
+\int \int \int (
a(a-1) x^{a-2}y^bz^c  (\frac{(\partial_x n(x, y, z, t)^2)}{4n(x, y, z, t)} + \frac{L_x(x, y,z,t)^2}{n(x, y, z, t)})
 \\+ 2ba x^{a-1}y^{b-1}z^c  (\frac{(\partial_x n(x, y, z, t)\partial_y n(x, y, z, t))}{4n(x, y, z, t)} + \frac{L_x(x, y,z,t)L_y(x, y,z,t)}{n(x, y, z, t)})\\
+ 2ca x^{a-1}y^bz^{c-1}  (\frac{(\partial_x n(x, y, z, t)\partial_z n(x, y, z, t))}{4n(x, y, z, t)} + \frac{L_x(x, y,z,t)L_z(x, y,z,t)}{n(x, y, z, t)})
\\+ b(b-1) y^{b-2}x^az^c(\frac{(\partial_y n(x, y, z, t)^2)}{4n(x, y, z, t)} + \frac{L_y(x, y,z,t)^2}{n(x, y, z, t)})
\\+ 2cb y^{b-1}x^az^{c-1} (\frac{(\partial_z n(x, y, z, t)\partial_y n(x, y, z, t))}{4n(x, y, z, t)} + \frac{L_z(x, y,z,t)L_y(x, y,z,t)}{n(x, y, z, t)})\\
+ c(c-1) z^{c-2}x^ay^b(\frac{(\partial_z n(x, y, z, t)^2)}{4n(x, y, z, t)} + \frac{L_z(x, y,z,t)^2}{n(x, y, z, t)})
)dx dy dz\\
- a(a-1) (
\frac{(a-2)(a-3)}{4} \braket{x^{a-4}y^bz^c} 
+ \frac{b(b-1)}{2} \braket{y^{b-2}x^{a-2}z^c} 
+ \frac{c(c-1)}{2} \braket{z^{c-2}x^{a-2}y^b} 
) \\
-b(b-1) (
\frac{(b-2)(b-3)}{4} \braket{y^{b-4}x^az^c}
+ \frac{c(c-1)}{2} \braket{z^{c-2}x^ay^{b-2}} 
) \\
-\frac{c(c-1)(c-2)(c-3)}{4} \braket{z^{c-4}x^ay^b}.\\
\end{multline}
\noindent
In practical calculations, many terms in the above equation vanish for the moments of low orders. These analytical expressions will be used later for a numerical demonstration of this moment propagation theory in Sections \ref{sec:numdemo} and \ref{sec:numdemo2}.


\subsection{Method of Electron Potential Energy Surface}
\label{sec:epes}
The above theoretical formulation for quantum dynamics in terms of the moments is exact in principle, but the numerical evaluation of the time derivatives of the moments is highly complicated.
When/if the principle of the energy conservation holds for individual single-particle wave functions (which is generally not the case for TD-KS orbitals in RT-TDDFT simulation), one can derive an alternative simpler expression.
Although it is not universally applicable, such an idea of designing an effective potential energy for quantum-mechanical particles (in a many-particle system) has been explored in the literature \cite{Ando_2018, https://doi.org/10.1002/qua.24820}.
The energy of a quantum mechanical particle, $\epsilon=\braket{\hat{H}}$, can be generally expressed in terms of the moments such that
\begin{equation}\
\label{eq:ene}
\epsilon( \mathbf{x}, \dot{\mathbf{x}}) = \epsilon( \{x_i, \dot{x}_i\})
\end{equation}
where we used the short-hand notation $x_i \equiv \braket{x^i}$ and  the integer index $i$ refers to the order of the moment.  In principle, the integer $i$ goes from 1 to $\infty$, but the dimension of $\mathbf{x}$ would be finite in practice. 
Eq. \ref{eq:ene} generally holds despite its unknown analytical form because the single-particle wave function can be obtained from the moments and their first-order time derivatives as shown in Section \ref{sub:spwf-moment}.
Assuming that this single-particle energy is conserved (i.e. $
\dot{\epsilon}(\mathbf{x}, \dot{\mathbf{x}}) = 0$), 
we here examine how the second-order time derivative of the moments can be derived, using a one-dimensional case for simplicity. Using the chain rule on the variables in $\epsilon$ yields
\begin{equation}
\label{eq:firstexpansion}
\sum_{k} (\frac{\partial \epsilon(\mathbf{x}, \dot{\mathbf{x}})}{ \partial x_k} \dot{x}_k + \frac{\partial \epsilon(\mathbf{x}, \dot{\mathbf{x}})}{ \partial \dot{x}_k} \Ddot{x}_k )=0.
\end{equation}

\noindent
In certain cases, like the harmonic oscillator model we discuss later, the second-order time derivative for a particular order $k$, $\Ddot{x}_k$, depends only on the terms for which the energy derivative is taken with respect to $x_k$ and $\dot{x}_k$.  
Then, the second-order time derivative can be solved for each order $k$ using
\begin{equation}
\frac{\partial \epsilon(\mathbf{x}, \dot{\mathbf{x}})}{ \partial x_k} \dot{x}_k + \frac{\partial \epsilon(\mathbf{x}, \dot{\mathbf{x}})}{ \partial \dot{x}_k} \Ddot{x}_k =0.
\end{equation}
Thus, the expression for the second-order time derivative is
\begin{equation}
\label{eq:EPES}
\Ddot{x}_k \equiv \frac{d^2 \braket{x^k}}{d t^2}=  -\frac{\partial \epsilon(\mathbf{x}, \dot{\mathbf{x}})}{ \partial x_k} \left(\frac{\partial \epsilon(\mathbf{x}, \dot{\mathbf{x}})}{ \partial \dot{x}_k}\right)^{-1} \dot{x}_k .
\end{equation}

\noindent
According to this Electron Potential Energy Surface (EPES) expression, the second-order time derivatives can be also evaluated from the energy $\epsilon$ by expressing it in terms of the moments and their first-order time derivatives. 
When the single-particle energy does not change in time, this EPES approach (Eq. \ref{eq:EPES}) is numerically more convenient compared to the more complicated expression, given by Eq. \ref{eq:secondderMPT2up}.
Eq. \ref{eq:EPES} shows some similarity to the expression in classical mechanics, where the second-order time derivative of the position (i.e. acceleration) is proportional to the spatial derivative of the potential energy. However, here, quantum effects are accounted for by an extra term, which would be the unity in classical mechanics analogue.


\subsection{Numerical Propagation of Moments}
Formulating quantum dynamics as the time propagation of the moments offers a convenient molecular dynamics (MD)-like scheme without explicitly integrating the TD-SE. The moment propagation becomes particularly attractive when the single-particle wave functions are spatially localized, as for those in the MLWF or simply the Wannier gauge in RT-TDDFT \cite{10.1063/1.5095631,doi:10.1063/5.0057587}.
In principle, infinite orders of the moments are necessary if one wishes to reproduce the exact quantum dynamics by propagating the moments. 
In practice, however, essential aspects of quantum dynamics relevant to calculating many physical properties can be reproduced with the moments of low orders. 
For example, the frequency-dependent dielectric function can be obtained from the dynamics of the first-order moment of single-particle wave functions in RT-TDDFT\cite{doi:10.1063/5.0057587}, and we anticipate that only a few orders of the moments need to be propagated for accurately modeling the dynamics of the first-order moment. 

Here, we propose a practical method for performing a numerical simulation using the moment propagation theory developed above. We restrict our discussion here to a one-dimensional case, for simplicity.  As done in classical MD simulation, the moments at a future time (i.e. $+\Delta t$ ) are Taylor expanded to the 2nd order,
\begin{equation}
\label{eq:taylor_moments}
\braket{x^a}(t+\Delta t)=\braket{x^a}(t)
+\Delta t \frac{d \braket{x^a}(t)}{d t} 
+\frac{1}{2}\Delta t^2 \frac{d^2 \braket{x^a}(t)}{d t^2}
+O(\Delta t^3 )
\end{equation}
where the time-step $\Delta t$ can be taken arbitrarily small so that $O(\Delta t^3 )$ is negligible. 
In order to employ a numerical integrator for using Eq. \ref{eq:taylor_moments}, the time derivatives of the moments need to be evaluated.
For many numerical integrators like Verlet algorithm for classical MD simulation, 
an analytical expression is necessary only for the second-order time derivatives. When the first-order time derivatives are also needed as in the velocity Verlet algorithm, they can be calculated numerically from the second-order time derivatives, etc. 
In classical mechanics, the second-order time derivative is proportional to the gradient of the potential energy via Newton's second law and it is a function of the position of atoms.
For the quantum dynamics, the moment propagation theory gives the necessary expression for the second-order time derivatives of the moments in terms of the moments and their first-order time derivatives. The second-order time derivatives are evaluated using Eq. \ref{eq:secondderMPT2up}. 
Alternatively, the EPES expression (Eq. \ref{eq:EPES}) can be used if the single-particle energy is conserved in the dynamics. 


\subsection{Analytical Solution for Simple Harmonic System}
\label{subsec:analytical}
In order to demonstrate the moment propagation theory and examine its feasibility using numerical methods in the following section, we use a simple one-dimensional harmonic potential, $V(x) = x^2$, for which we can derive the exact analytical solution. We consider a time-dependent Hamiltonian such that simple homogeneous electric field is applied to the system. 
While this might be considered a oversimplified model, note that the dynamics of individual single-particle wave functions, localized on chemical bonds, are often confined in such a localized potential well in the MLWF gauge\cite{10.1063/1.5095631}.
We restrict ourselves to the situation in which the quantum particle is initially in the equilibrium ground state, and thus only the second-order Edgeworth (E2) expansion (i.e. $n$=2 for Eq. \ref{eq:pdf}) is needed for deriving the analytical equation of motion \cite{Andrews_2016, Blinnikov_1998, 10.2307/2237255}. The particle density is given by 

\begin{equation}
n(x, t) = \frac{1}{\sqrt{2 \pi (\braket{x^2}(t)- (\braket{x}(t))^2)} } \exp{(-\frac{1}{2} \frac{(x - \braket{x}(t))^2}{\braket{x^2}(t)- (\braket{x}(t))^2})}.
\end{equation}

\noindent
As a commonly encountered situation in which the light is treated as a classical electromagnetic field, a spatially homogeneous electric field is used as the time-dependent external potential that perturbs the quantum system such that $V(x, t)= x^2 +c(t)x$ in the length gauge. 
One can find an exact analytical solution for the second-order time derivatives using the above described theory of moment propagation, and the second-order time derivative of the first-order moment (see Appendix \ref{sec:eq27derivation}) is

\begin{equation}
\label{eq:FMTD}
\frac{d^2 \braket{x}(t)}{d t^2}= -2\braket{x}(t)-c(t)
\end{equation}
and, for the second-order moment, we have
\begin{equation}
\label{eq:SMTD}
\frac{d^2 \braket{x^2}(t)}{d t^2}= -4 \braket{x^2}(t) -2c(t)\braket{x}(t) + 2((\frac{d\braket{x}(t)}{d t})^2 + \frac{1 + (\frac{d S(t)}{d t})^2}{4 S(t)}) 
\end{equation}
where $S(t)= \braket{x^2}(t) - \left(\braket{x}(t)\right)^2$ is essentially the spread or the variance of the corresponding single-particle wave function. 
For the harmonic oscillator with a homogeneous external electric field, only the two lowest orders of the moments are necessary to describe the quantum dynamics exactly.
The energy of the quantum particle, which is equivalent to $\epsilon$ for this one-particle system, can be written (see Appendix \ref{sec:eq27derivation}) as 
\begin{equation}
\label{eq:EneTh}
E(t) = \epsilon(t)= \braket{x^2}(t)+c(t)\braket{x}(t)+\frac{1}{2}((\frac{d \braket{x}(t)}{d t})^2 + \frac{1 + (\frac{d S(t)}{d t})^2}{4 S(t)}).
\end{equation}

\noindent
In the context of EPES method discussed in Section \ref{sec:epes}, one could alternatively use this analytical energy expression (Eq. \ref{eq:EneTh}) to derive the second-order time derivatives of the moments as shown in Appendix \ref{sec:eq27derivation}. As expected, the EPES method gives the exact same analytical solutions for the second-order time derivatives of the moments (i.e. Eq. \ref{eq:FMTD} and \ref{eq:SMTD}).

\section{Numerical Demonstration}

\subsection{Harmonic Oscillator Potential}

\label{sec:numdemo}
A simple one-dimensional system is used here to demonstrate the numerical methods using the moment propagation theory described in the above sections. 
As a proof-of-principle demonstration, we consider $V(x, t)= x^2 +c(t)x$. This represents a simple harmonic oscillator with time-dependent external electric field applied to the system using the length gauge in Hamiltonian. 
We employ spatially homogeneous electric field such that
\begin{equation}
\label{eq:ct}
c(t)=
    \begin{cases}
        0.3 \text{ a.u.} &  0 \text{ a.u.} < t < 0.45 \text{ a.u.}\\
        0 \text{ a.u.}&  \text{else.}
    \end{cases}
\end{equation}
\noindent
The electric field is applied initially to perturb the system for the first 449 time-steps (0.45 a.u.), and the system is propagated for a total of 12001 time steps (12 a.u.). 
For this harmonic potential with the time-dependent electric field, the analytical solution can be derived as discussed in Subsection \ref{subsec:analytical} as the reference for examining the numerical methods. 

For the moment propagation theory (MPT)-based methods, 32 real-space grid points are used as the basis over the spatial range of 7 a.u., and 
Eq. \ref{eq:taylor_moments} was used for the explicit time integration with a time-step of $\Delta t = $ 0.001 a.u. using the Newmark-beta method, with parameters $\gamma = \frac{1}{2}$ and $\beta = \frac{1}{8}$ \cite{doi:10.1061/JMCEA3.0000098}. 
For the MPT methods, we show the simulation results using the second-order Edgeworth approximation (MPT-E2) and the fourth-order Edgeworth approximation (MPT-E4). 
Here, the order, $n=2,4$, in the Edgeworth approximation (i.e. E2, E4), corresponds the $n$ in Eq. \ref{eq:pdfn}.
The second-order time derivatives are calculated explicitly using Eq. \ref{eq:secondderMPT2up}. 
Additionally, we performed the MPT simulation by employing the EPES approach, in which the the second-order time derivatives are calculated from the energy according to Eq. \ref{eq:EPES}. The energy here is expressed as (see Appendix \ref{sec:eq31derivation})  

\begin{equation}
\label{eq:En}
E = \int V(x) n(x) dx +\frac{1}{2} \int   (\frac{(\frac{\partial n(x)}{\partial x}^2)}{4n(x)} + \frac{L(x)^2}{n(x)}) dx
\end{equation}
where $L(x)$ is given by Eq. \ref{eq:Lrup}. In order to evaluate the particle density here, the second-order Edgeworth approximation is employed, and we refer to this approach as MPT-EPES-E2. 
For comparison, we also perform the standard propagation of the wave function by directly integrating the time-dependent Schrodinger equation, which we refer to as wave function theory (WFT). For the WFT method, the wave function is represented using the real space grids with 100 evenly spaced points over the same 7 a.u. range as for the MPT methods. The Crank-Nicholson method was used to numerically integrate the wave function with a time-step of $\Delta t = $ 0.001 a.u.\cite{crank_nicolson_1947}. In the limit of infinitely dense grid points and an infinitely small time-step, all the methods converge to the exact analytical result for this particular harmonic potential case.  

\begin{figure}[h]
\captionsetup{justification=centerlast}

    \begin{subfigure}{0.48\textwidth}
    \includegraphics[width=0.9\linewidth]{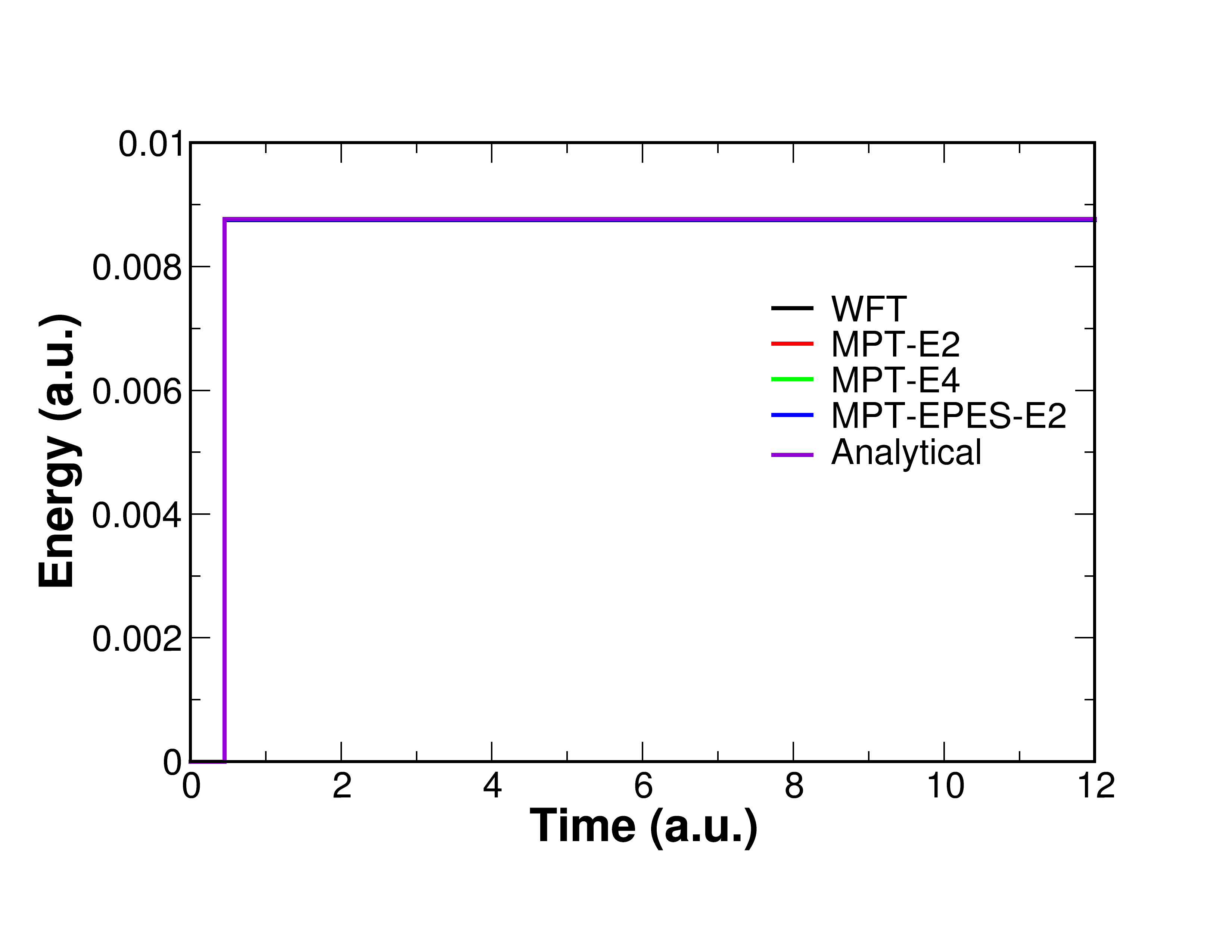}
    \caption{} 
    \label{fig:sub0im1}
    \end{subfigure}
    \begin{subfigure}{0.48\textwidth}
    \includegraphics[width=0.9\linewidth]{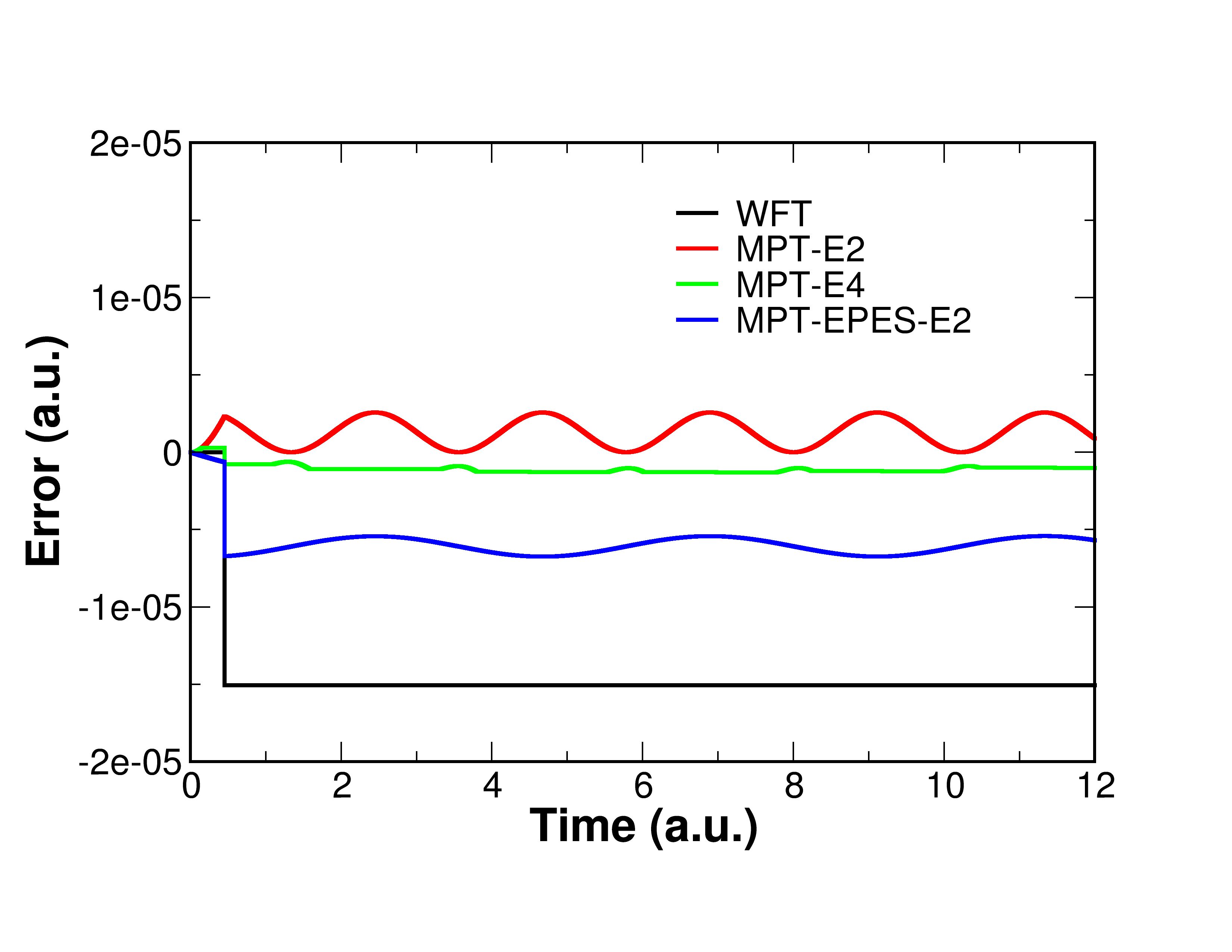}
    \caption{}
    \label{fig:sub0im2}
    \end{subfigure}
    
    \caption{
    (a) Total energy as a function of time for the  exactly-solvable one-dimensional (1D) harmonic potential model, $V(x,t) = x^2+c(t)x$. The homogeneous electric field is applied for t=0-0.45 a.u., and the reference energy (E=0.0 a.u.) is set to that of the initial ground state at t=0 a.u.. The three MPT methods and the standard wave function theory (WFT) method are compared to the exact analytical solution. (b) The error on the total energy with respect to the exact analytical solution. See the main text for more details.} 
    
    \label{fig:energy}
\end{figure}

\begin{figure}[h]
\captionsetup{justification=centerlast}

    \begin{subfigure}{0.48\textwidth}
    \includegraphics[width=0.9\linewidth]{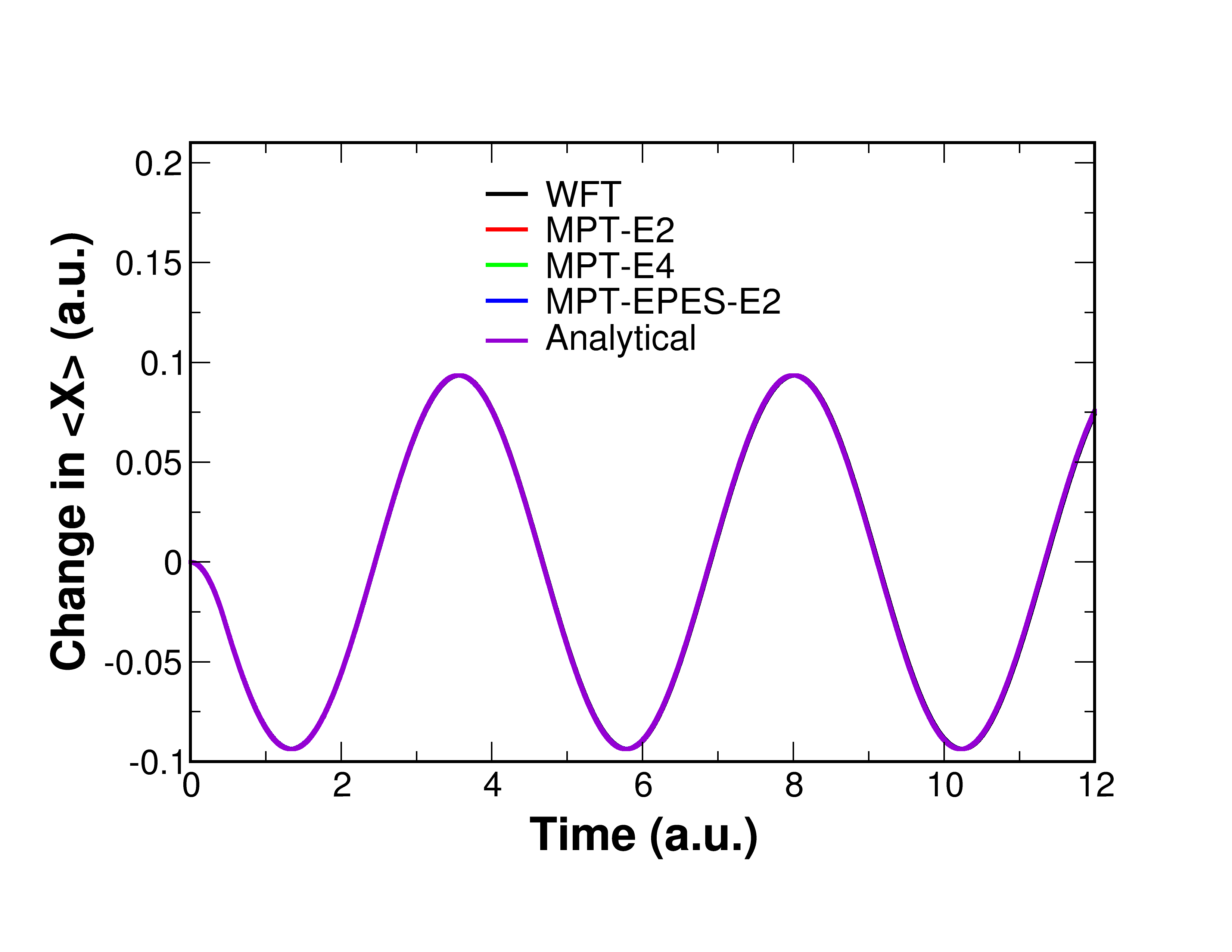}
    \caption{} 
    \label{fig:subim1}
    \end{subfigure}
    \begin{subfigure}{0.48\textwidth}
    \includegraphics[width=0.9\linewidth]{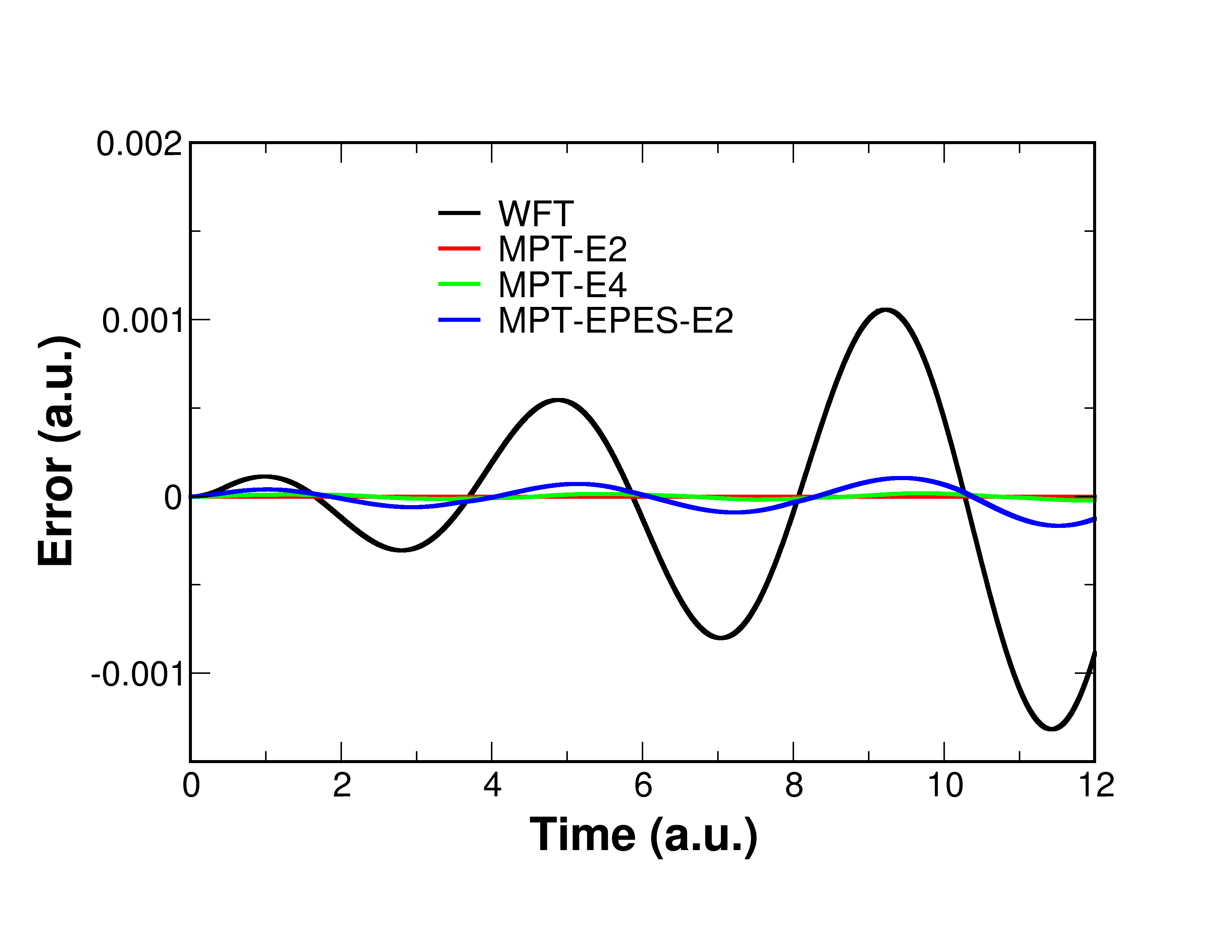}
    \caption{}
    \label{fig:subim2}
    \end{subfigure}
    
    \caption{
    Comparison of the three MPT methods and the standard wave function theory (WFT) method to the analytical solution for (a) the first-order moment as a function of time and (b) its error with respect to the exact analytical solution. See also Figure \ref{fig:energy} for additional details.} 
    \label{fig:xmoment}
\end{figure}
Figure \ref{fig:energy} shows the total energy, $E$, as a function of time, and the energy is expected to remain constant after the perturbing electric field is turned off.
Having the exact analytical solution for this system, we know that the energy change must be $\Delta E = 8.771 \times 10^{-3}$ a.u. in principle if all numerical parameters are fully converged and computations are performed exactly. 
With the real-space basis of 32 grid points, the MPT methods (MPT-E2 and MPT-E4) yield $\Delta E$ values of $8.772 \times 10^{-3}$ a.u. and $8.770 \times 10^{-3}$ a.u., respectively. 
The MPT method with the EPES approach (MPT-EPES-E2) gives a slightly smaller value of $\Delta E$ of $8.766 \times 10^{-3}$ a.u.. Compared to the MPT-E2 method, the MPT-EPES-E2 method gives a larger error because of the numerical evaluation of the second-order time derivative in Eq. \ref{eq:EPES} through the partial derivative of the energy with respect to the moments and their time derivatives. 
With a larger real-space basis of 100 grid points, the standard approach of propagating the wave function (WFT) gives a similar value of $\Delta E$=$8.756 \times 10^{-3}$ a.u.. While the WFT method gives the largest error for this particular harmonic oscillator potential, all the numerical methods show sufficient accuracy in terms of the relative error on the energy. 
Table \ref{table:table1} 
also summarizes the energy conservation for all methods, and all numerical methods show that the energy drift/deviation is at least several orders of magnitude smaller than the energy change due to the external electric field. 

\begin{table}[!ht]
\captionsetup{justification = centerlast}
    \centering
    \begin{tabularx}{\columnwidth}{X>{\hsize=.5\hsize}X>{\hsize=.5\hsize}X>{\hsize=.5\hsize}X>{\hsize=.5\hsize}X}
    \hline\hline
        Method & WFT & MPT-E2 & MPT-E4 & MPT-EPES-E2 \\ \hline
        Energy Change, $\Delta E$ (a.u.) & $8.756 \times 10^{-3}$ & $8.772 \times 10^{-3}$ & $8.770 \times 10^{-3}$ & $8.766 \times 10^{-3}$ \\ 
        Energy Drift (a.u.) per 1 a.u. time  & $1.150 \times 10^{-12}$ & $1.240 \times 10^{-7}$ & $2.073 \times 10^{-8}$ & $8.955 \times 10^{-8}$  \\ 
        Oscillation Frequency (a.u.) & $1.41296$ & $1.41421$ & $1.41420$ & $1.41409$  \\ \hline \hline
    \end{tabularx}
    \caption{ Energy change, energy drift per 1 a.u. time, and the oscillation frequency of the first-order moment for the simulation shown in Figure \ref{fig:energy}. 
    The exact analytical value of $\Delta E$ is $8.771 \times 10^{-3}$ a.u., and the exact analytical value for the oscillation frequency is $\sqrt2=1.41421$.} 
 
    \label{table:table1}
\end{table}

The first-order moments are of particular importance in quantum dynamics because they can be directly related to physical properties such as the frequency-dependent polarizability/conductivity, from which, for instance, the optical absorption spectrum can be calculated in electronic structure theory \cite{https://doi.org/10.1002/pssb.200642005, PhysRevA.21.1561}.  
As expected, the first-order moment of this quantum particle oscillates similarly to classical particle for this simple harmonic potential. 
Figure \ref{fig:xmoment} shows the comparison for the time evolution of the first-order moment among the newly-developed MPT methods and the standard WFT approach. 
Although the basis set size necessary for convergence is system-dependent in general, the MPT methods exhibit a smaller numerical error with only 32 grid points while the WFT simulation here used 100 grid points. 
Note that the differences among the numerical methods are rather negligible (see Fig. \ref{fig:subim1} ),
and they all show only small errors with respect to the exact analytical result,  
as seen in Figure \ref{fig:subim2}.
The oscillation frequency of the first-order moment, which would be related to the polarizability/conductivity in electronic structure calculations, does not deviate appreciatively from the exact analytical value of $\sqrt{2}$ for any of the numerical methods as shown in Table 
\ref{table:table1}.

\begin{figure}[h]
\captionsetup{justification=centerlast}

    \begin{subfigure}{0.48\textwidth}
    \includegraphics[width=0.9\linewidth]{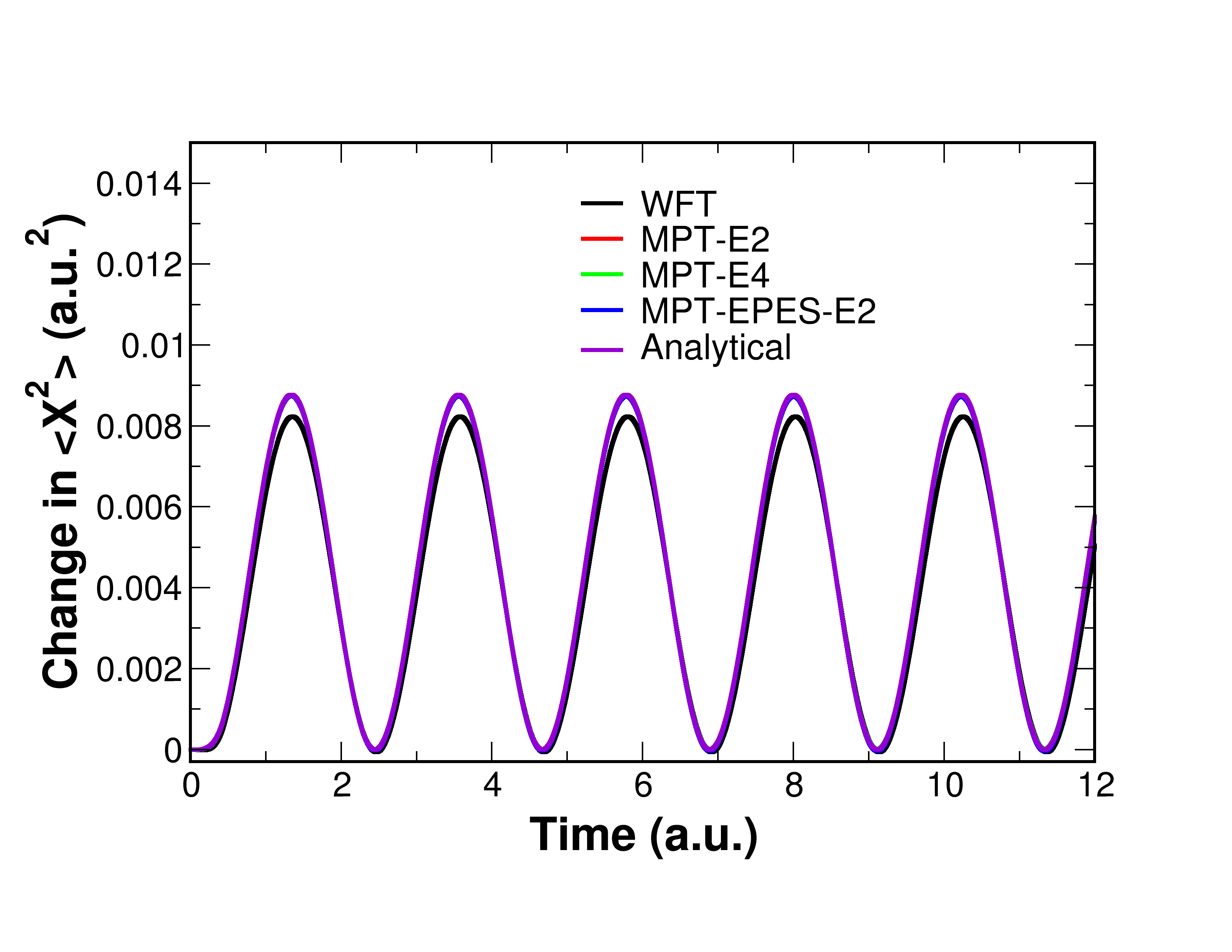}
    \caption{}
    \label{fig:sub2im1}
    \end{subfigure}
    \begin{subfigure}{0.48\textwidth}
    \includegraphics[width=0.9\linewidth]{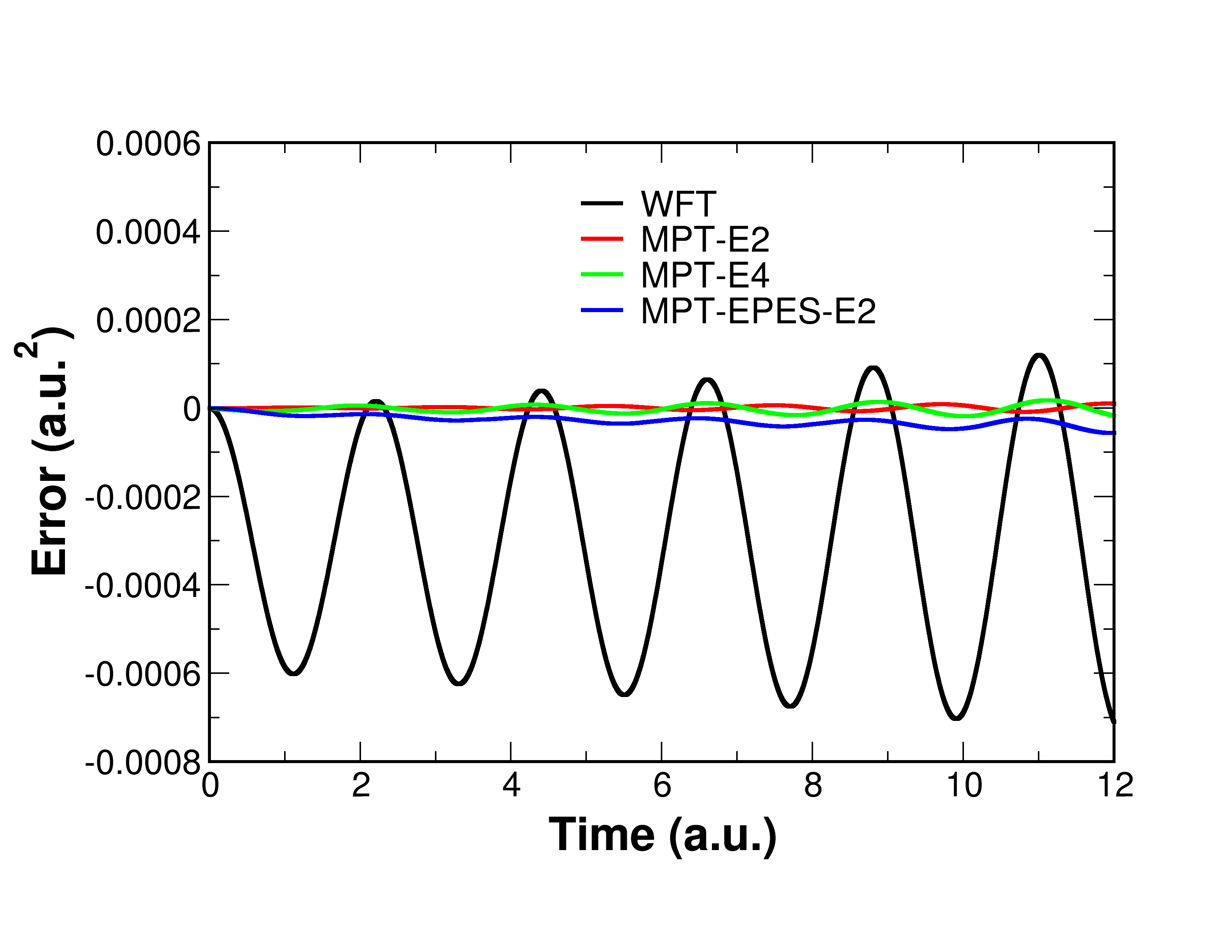}
    \caption{}
    \label{fig:sub2im2}
    \end{subfigure}
    
    \caption{
    Comparison of the three MPT methods and the standard wave function theory (WFT) method to the analytical solution for (a) the second-order moment as a function of time and (b) its error with respect to the exact analytical solution. See also Figure \ref{fig:energy} for additional details.} 
    \label{fig:x2moment}
\end{figure}

Figure \ref{fig:x2moment} shows the time evolution of the second-order moment according to the different numerical methods, along with the errors with respect to the exact analytical solution. 
The second-order moment is related to the spatial variance (i.e. the spread) of the quantum particle.
In this particular case, all the MPT methods show much smaller errors than the WFT method although  all the errors are sufficiently small (see Fig. \ref{fig:sub2im2}). 
Comparing to the exact analytical solution, these results numerically confirm the validity of the newly-developed MPT approach and the practical feasibility of employing numerical methods based on this new theoretical formulation. As demonstrated here, the MPT numerical methods perform as well as the WFT method, if not better, for this simple case with the analytical solution available. 

\subsection{Anharmonic Morse Potential}
\label{sec:numdemo2}

\begin{figure}[h]
\captionsetup{justification=centerlast}

    \begin{subfigure}{0.48\textwidth}
    \includegraphics[trim=0 0 0 0,clip,width=0.9\linewidth]{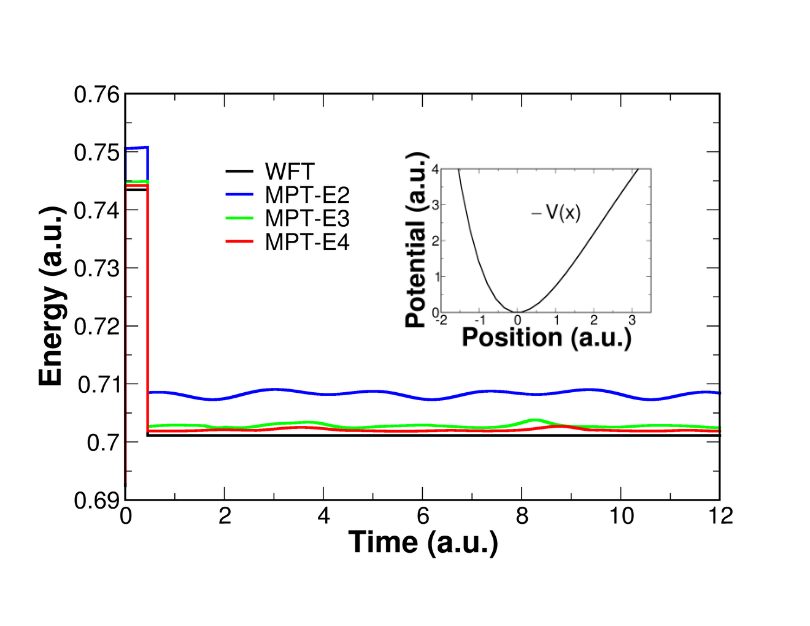}
    \caption{}
    \label{fig:sub3im1}
    \end{subfigure}
    \begin{subfigure}{0.48\textwidth}
    \includegraphics[width=0.9\linewidth]{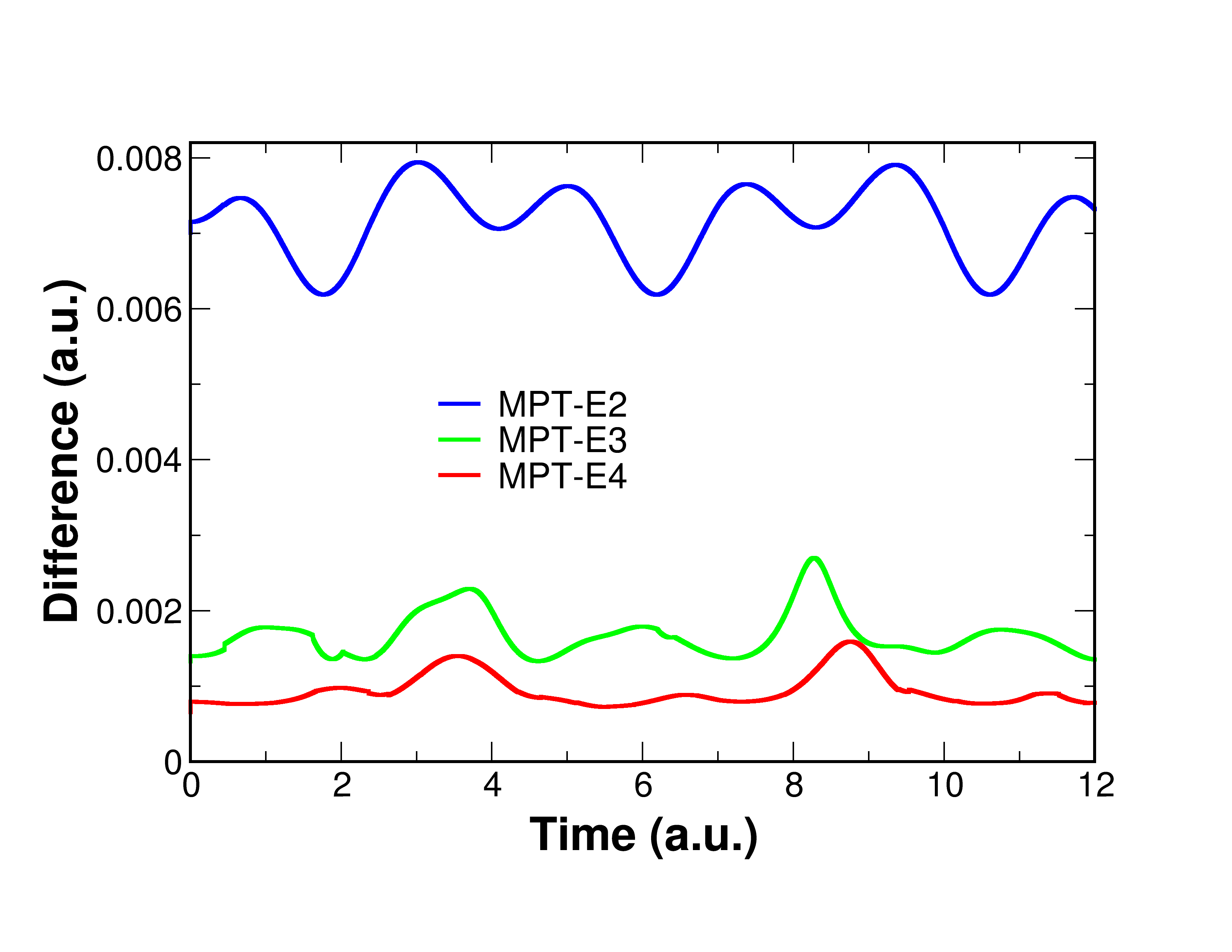}
    \caption{}
    \label{fig:sub3im2}
    \end{subfigure}
    
    \caption{
    (a) Total energy as a function of time for the anharmonic 1D model with Morse potential (Eq. \ref{eq:morse}). The inset shows the Morse potential energy function $V(x)$. The homogeneous electric field is applied for t=0-0.45 a.u.
    The MPT method with an increasing order of the moments is compared to the standard wave function theory (WFT) method.  (b) The differences with respect to the WFT method result are shown. }
    \label{fig:energya}
\end{figure}

Although an exact analytical solution is not available for this anharmonic potential, we also examined the performance of the numerical MPT method for the Morse potential,

\begin{equation}
\label{eq:morse}
V(x,t) = a(1-e^{-\sqrt{\frac{1}{a}} x})^2+c(t)x
\end{equation}
where $a$ determines the anharmonicity of this Morse potential, and we set $a=10$ a.u. (see Fig. \ref{fig:energya}). 
The same external electric field, given by Eq. \ref{eq:ct}, is applied to the system through the $c(t)x$ term in the length gauge. All computational parameters are kept the same as the harmonic oscillator potential case discussed above, except for the grid size of 6 a.u..
Figure \ref{fig:energya} shows the total energy as a function of time for the MPT method with increasingly higher orders of moments, along with the standard WFT method for comparison. 
Although an exact analytical solution is not available for this model, the energy must be conserved in a numerically exact propagation. Figure \ref{fig:energya}(a) shows better energy conservation during the dynamics with less fluctuations for the MPT method with increasingly higher orders of the moments.  
In this particular case, the WFT method shows slightly better energy conservation than the MPT method even when up to the 4th-order moments are included (i.e. MPT-E4). Convergence of the MPT method and the standard WFT method can be seen in
Figure \ref{fig:energya}(b), as the Edgeworth expansion becomes more complete by including  higher orders of the moments. 

\begin{figure}[h]
\captionsetup{justification=centerlast}

    \begin{subfigure}{0.48\textwidth}
    \includegraphics[width=0.9\linewidth]{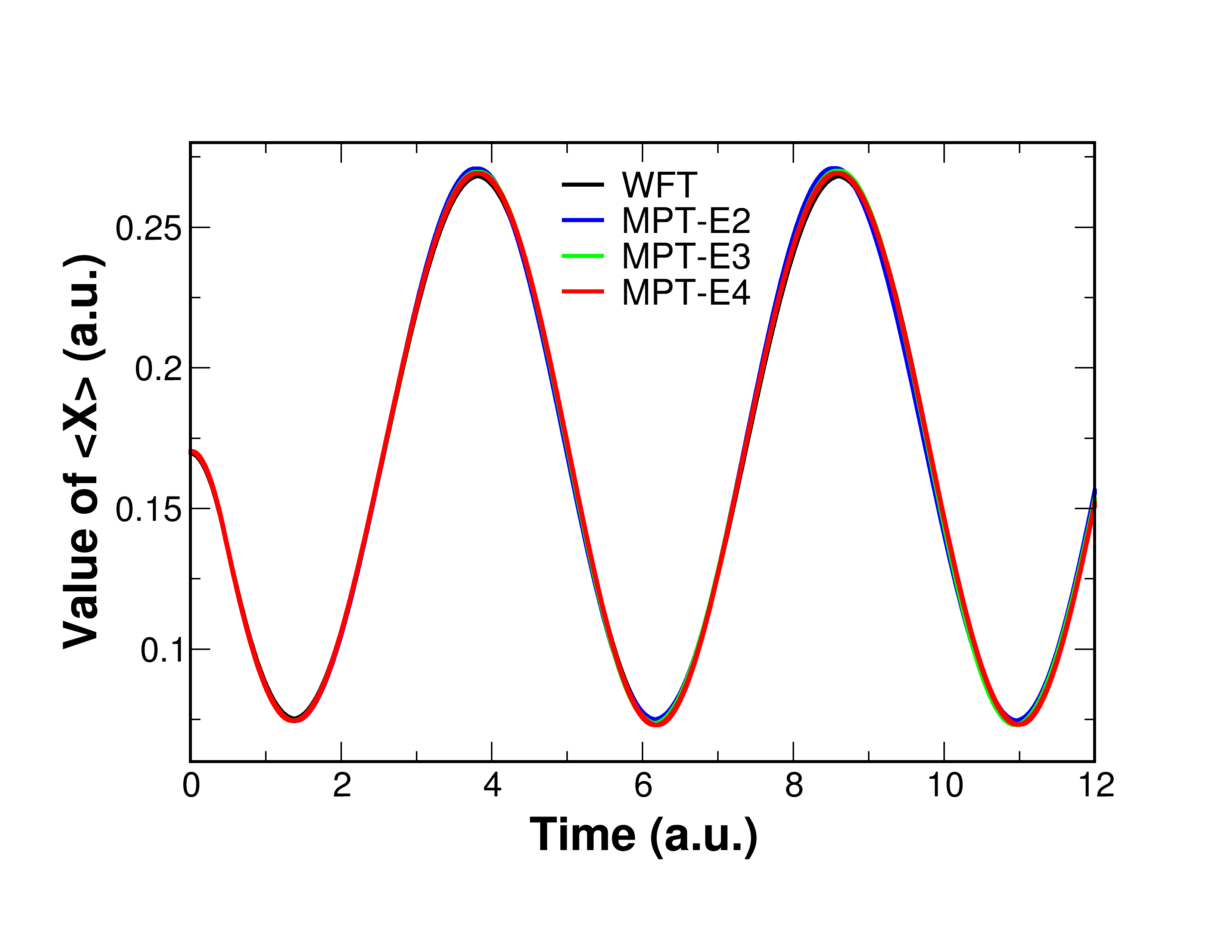}
    \caption{}
    \label{fig:sub3im1}
    \end{subfigure}
    \begin{subfigure}{0.48\textwidth}
    \includegraphics[width=0.9\linewidth]{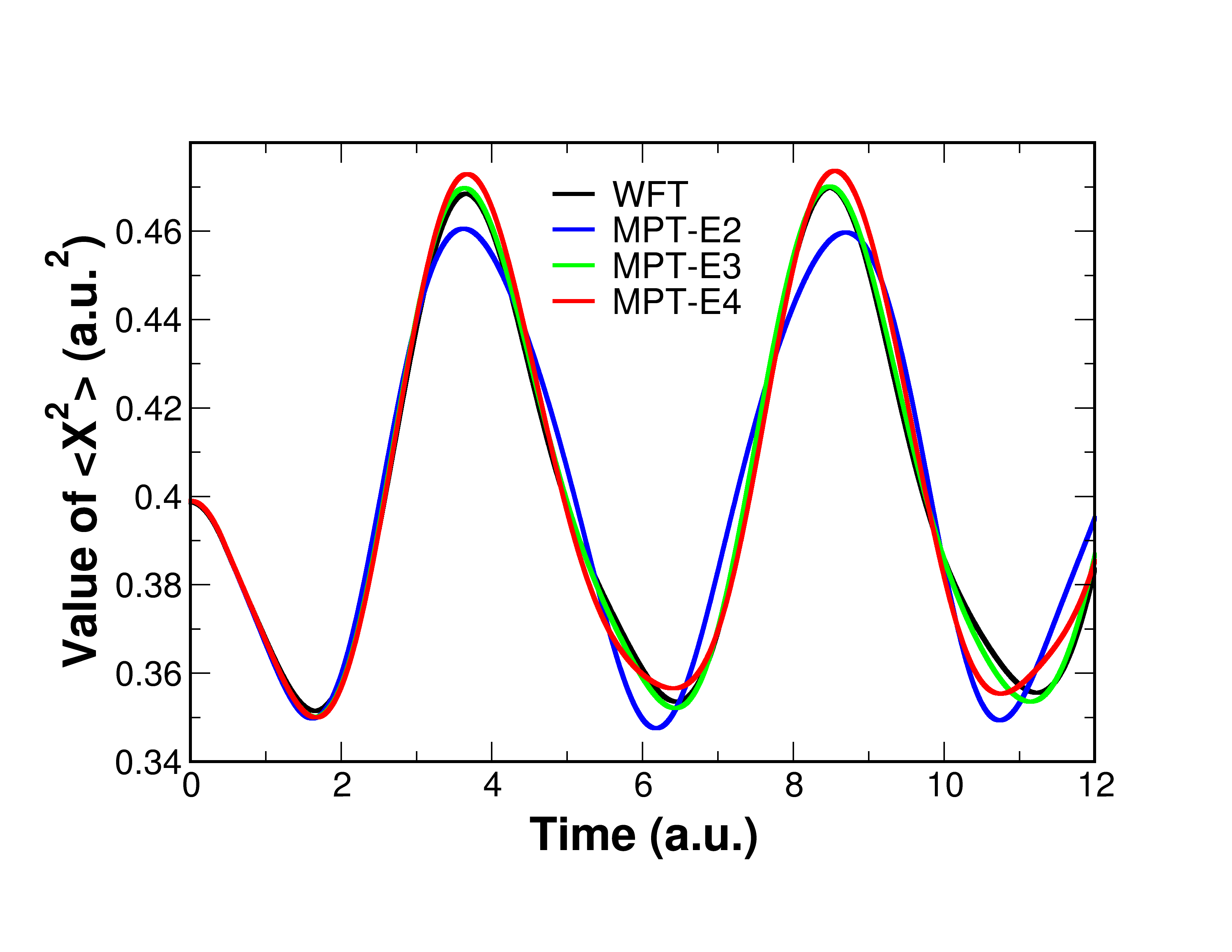}
    \caption{}
    \label{fig:sub3im2}
    \end{subfigure}
    
    \caption{
    (a) first-order moment and (b) second-order moment as a function of time for the anharmonic 1D model (Eq. \ref{eq:morse}). 
    The MPT method with an increasing order of the moments is compared to the standard wave function theory (WFT) method. See also Figure \ref{fig:energya} for additional details.} 
    \label{fig:xa}
\end{figure}

Figure \ref{fig:xa} shows the time evolution of the first-order and the second-order moments in the quantum dynamics. The MPT dynamics becomes increasingly closer to the conventional WFT method as the higher orders of the moments are included in the MPT method. 
This is particularly apparent for the second-order moment (see Fig. \ref{fig:xa}(b)) as the MPT-E4 result closely resembles the WFT method result while significant deviations can be seen for the MPT-E2 result.

Although the moment propagation theory (MPT) is an exact reformulation of quantum dynamics in terms of the moments instead of the single-particle wave function, its usefulness might not be obvious in the above proof-of-principle demonstrations. 
A key advantage of this novel MPT is that machine-learning techniques can be readily used for modeling the second-order time derivatives of the moments instead of explicitly calculating them. 
For classical MD simulations, machine-learning techniques, particularly modern neural network models, are widely used for evaluation of the potential energy and the atomic forces, and a similar approach can be conveniently adapted for quantum dynamics simulation using the MPT. 
This idea is briefly demonstrated in the following section. 


\section{Machine-Learning Time Derivatives with Artificial Neural Network}
\begin{figure}[!htp]
\captionsetup{justification = centerlast}

    \begin{subfigure}{0.48\textwidth}
    \includegraphics[width=0.9\linewidth]{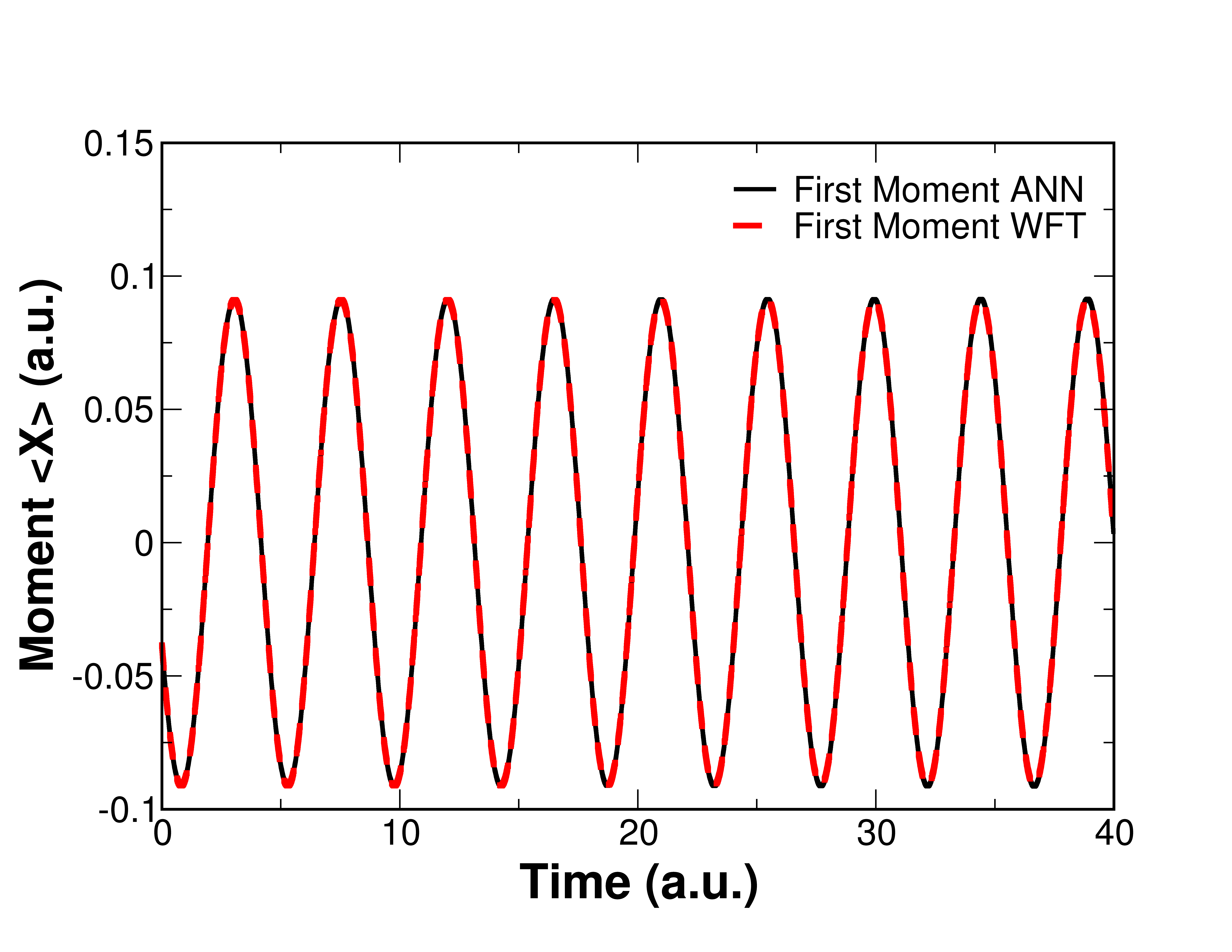}
    \caption{}
    \label{fig:sub3im1}
    \end{subfigure}
    \begin{subfigure}{0.48\textwidth}
    \includegraphics[width=0.9\linewidth]{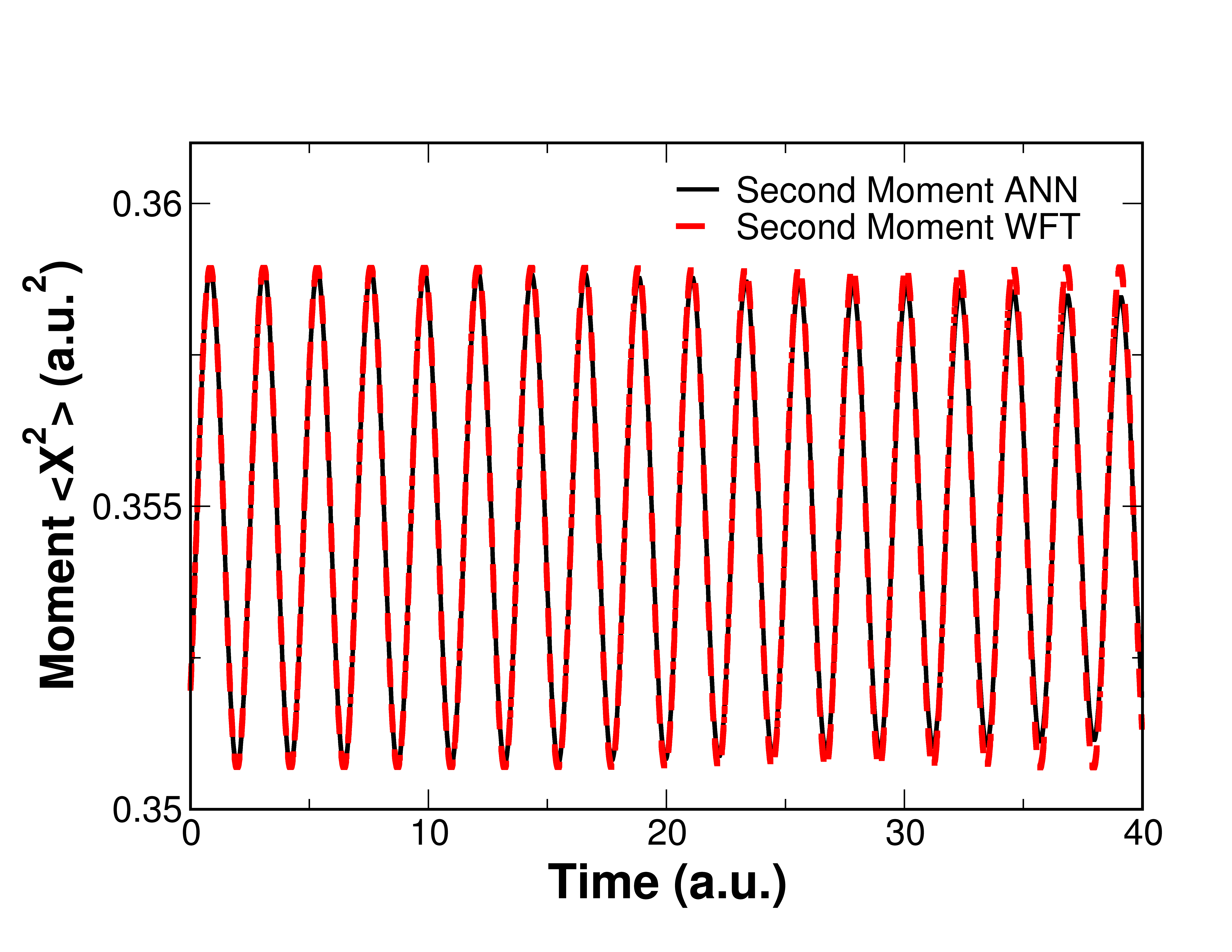}
    \caption{}
    \label{fig:sub3im2}
    \end{subfigure}
    
    \caption{
    The use of ANN model for the calculation of the second-order time derivative of the moments in the MPT method. 
   Comparison of (a) the first-order moment and (b) the second-order moment in the MPT simulation based on the ANN model and explicit WFT simulation. Based on the analytical solution for this exactly-solvable harmonic potential model, the input descriptors for the ANN model consist of $\braket{x}$, $\braket{x^2}$, $(\frac{\partial \braket{x}}{\partial t})^2$, and $(\frac{\partial S}{\partial t})^2$.  The WFT simulation was used also to provide the training data for the ANN model.}
    \label{fig:ANNtestx2}
\end{figure}

Simulating quantum dynamics is highly nontrivial and computationally expensive, and this often limits our ability to study long-time dynamics of quantum systems and/or statistical properties that derive from an ensemble of quantum dynamics trajectories.
There have been a number of works to employ modern machine-learning approaches like artificial neural network (ANN) models for efficiently performing such quantum dynamics simulations instead of numerically integrating the time-dependent Schrodinger equation \cite{yao2022emulating, doi:10.1021/acs.jpclett.1c03117, doi:10.1021/acs.jpclett.0c02168, doi:10.1021/acs.jctc.2c00702} 
In this section, we demonstrate how the newly-developed MPT offers an convenient avenue for developing an ANN model for quantum dynamics simulation. 
For classical MD simulations, the use of ANN has become widespread in recent years \cite{doi:10.1021/acs.chemrev.0c01111, doi:10.1021/acs.jpcc.6b10908, PhysRevLett.120.143001}.  
In particular, machine-learning techniques are used to construct an accurate potential energy function (i.e. force field) as an ANN through training with first-principles quantum-mechanical calculations \cite{10.1063/5.0012815, doi:10.1021/acs.jpclett.1c01566, Han_2018, doi:10.1073/pnas.1907975116, PhysRevLett.120.143001, PhysRevLett.98.146401}.
This allows for an efficient computation of the accurate force on atoms, which is proportional to the second-order time derivative of the classical particle positions. 
We here demonstrate a machine-learning of the second-order time derivative of the moments as an ANN model for performing the quantum dynamics simulation within the MPT framework.
In particular, we examine the construction of ANN models for the second-order time derivatives of the moments using $\braket{x}$, $\braket{x^2}$, $(\frac{\partial \braket{x}}{\partial t})^2$, and $(\frac{\partial S}{\partial t})^2$ as the input descriptors for the harmonic potential model. 
The choice of these descriptors are motivated by the analytical solution. 
A simple ANN model was used here, consisting of two hidden layers with eight Scaled Exponential Linear Units (SELU) activation nodes\cite{chollet2015keras}. The input descriptors are multiplied by weights and they acquire an added bias when moving to the next layer. 
The activation function, SELU, acts on each input for each node and its output is fed into the next layer. There are many different approaches for developing ANNs, but we chose this simple proof-of-concept architecture \cite{doi:10.1021/acs.chemrev.0c01111}. The mean squared error (MSE) was used for the loss function, and the ANN training used the Adam optimizer\cite{chollet2015keras}. 
Each ANN model was trained for at least a thousand epochs until the MSE reaches a minimum. The training data was generated from propagating the wave function according to the TD-SE with a time-step of $\Delta t = 0.01$ a.u..
For training each ANN model, a total of 9498 data points were used.
ANN model then replaces the numerical evaluation of the second-order time derivative of the moments using Eq. \ref{eq:secondderMPT2up} in the quantum dynamics simulation based on the MPT.
Figure \ref{fig:ANNtestx2} shows the first-order and second-order moments from the MPT simulation using the ANN model (MPT-ANN) for the harmonic potential, compared against its own training data as a proof-of-principle demonstration. 
The MPT-ANN simulation successfully reproduces the time evolution of the moments to a great accuracy. 
For the second-order moment,  a slight deviation is observed for longer times while the oscillation frequency remains accurately reproduced. Improving the machine-learning techniques for quantum dynamics simulation with the MPT will be a topic of a separate study in the future. 
The computational procedure/cost is numerically comparable to that of performing classical MD simulation using the ANN model for the force evaluation. 

\begin{figure}[H]
\captionsetup{justification = centerlast}

    \begin{subfigure}{0.48\textwidth}
    \includegraphics[width=0.9\linewidth]{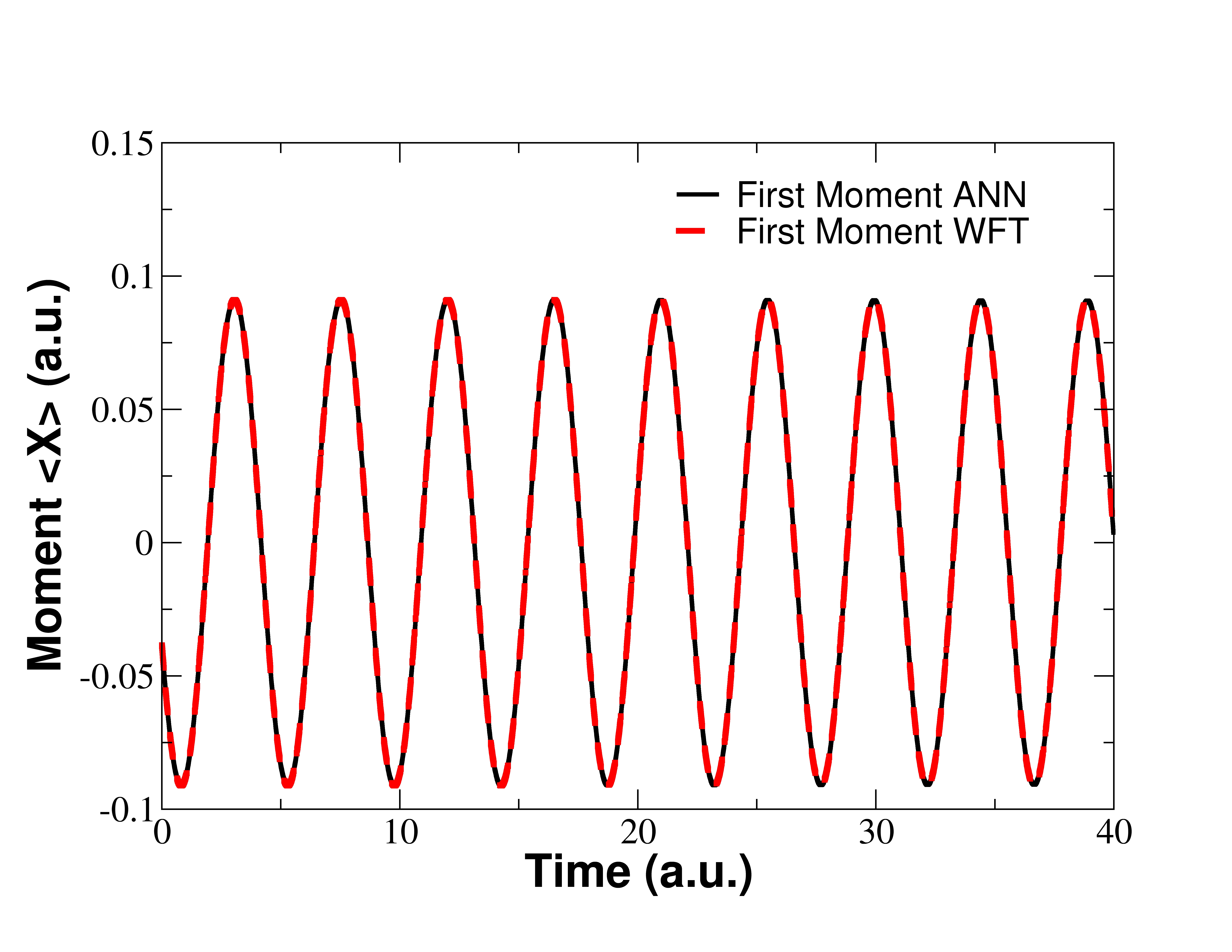}
    \caption{}
    \label{fig:sub4im1}
    \end{subfigure}
    \begin{subfigure}{0.48\textwidth}
    \includegraphics[width=0.9\linewidth]{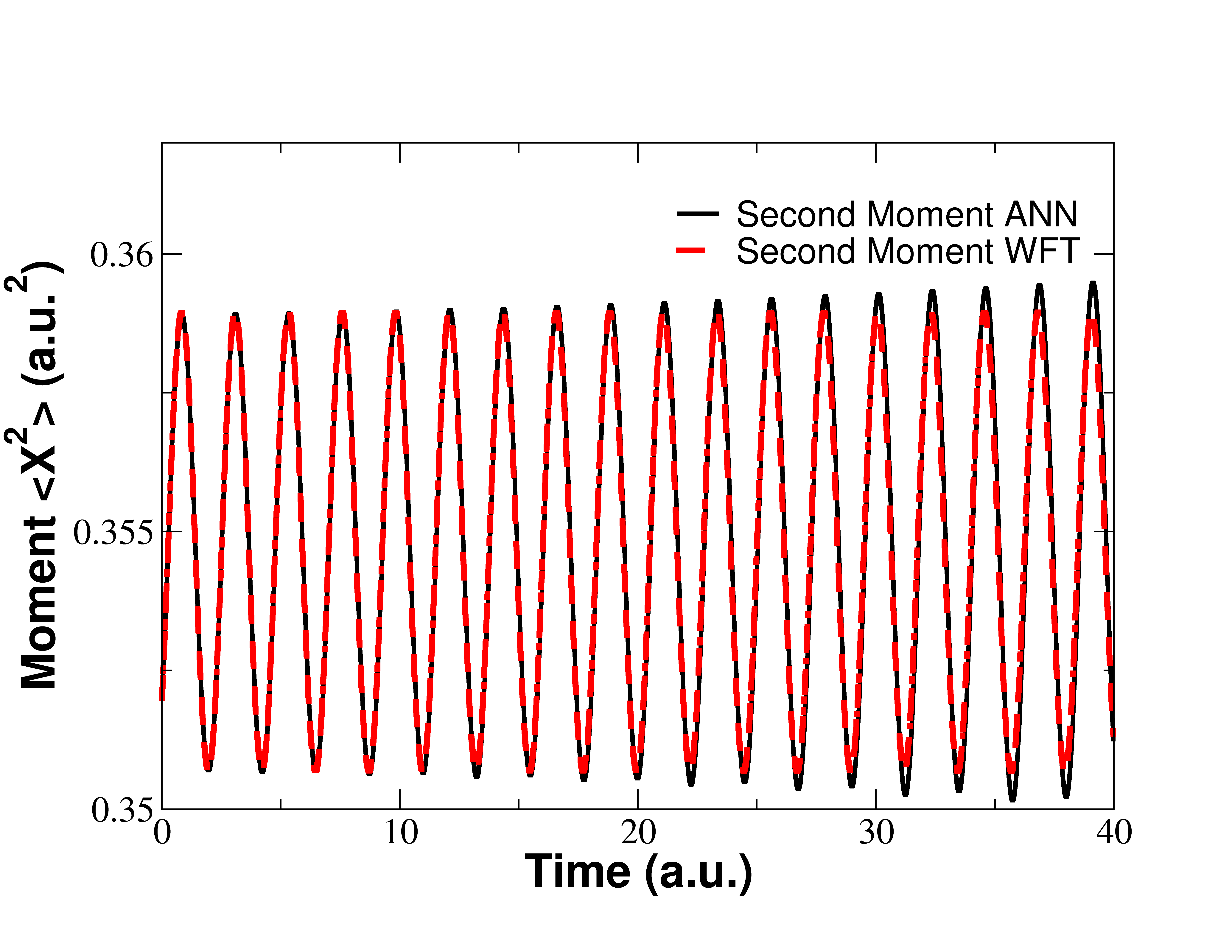}
    \caption{}
    \label{fig:sub4im2}
    \end{subfigure}
    
    \caption{
    The use of ANN model for the calculation of the second-order time derivative of the moments in the MPT method. Additional details are same as in Figure \ref{fig:ANNtestx2} but only $\braket{x}$ and $\braket{x^2}$ are used as the input descriptors for the ANN model. Comparison of (a) the first-order moment and (b) the second-order moment in the MPT simulation based on the ANN model and explicit WFT simulation. }
    \label{fig:ANNtestx3}
\end{figure}

We envision that machine-learning techniques like ANN models can be used to construct an accurate model of the second-order time derivative of the moments in general. 
For the harmonic potential model, the exact analytical expression can be derived and thus it was possible to deduce what physically-meaningful input descriptors were in terms of the moments for the ANN construction. 
However, in realistic first-principles electronic structure theory descriptions such as when electrons are represented by  time-dependent maximally-localized Wannier functions in RT-TDDFT simulation \cite{10.1063/1.5095631}, the potential that individual electrons are subjected to have no analytical solutions in general although confining in many cases. 
We briefly examine here how the quantum dynamics might be sensitive to the input descriptors in the ANN model using this exactly-solvable harmonic potential. 
Table \ref{table:RMSE} shows the oscillation frequency and the root mean squared error (RMSE) of the first-order moment as well as the RMSE of the second-order moment using different
input descriptors. 
Even with only $\braket{x}$ and $\braket{x^2}$ as the input descriptors for the ANN model, the RMSE remains quite small for both the first-order and second-order moments in the dynamics as shown in Figure \ref{fig:ANNtestx3}. 
Additionally, the oscillation frequencies from the MPT simulation using the ANN models with different input descriptors are essentially the same as the frequency obtained from the reference WFT simulation as seen in Table \ref{table:RMSE}.
In summary, we showed here how the newly-developed MPT provides 
a convenient theoretical framework for accelerating the quantum dynamics simulation through machine-learning the second-order time derivatives using a simple ANN model. 

\begin{table}[!ht]
\captionsetup{justification = centerlast}
    \centering
    \begin{tabularx}{\columnwidth}{X>{\hsize=.5\hsize}X>{\hsize=.5\hsize}X>{\hsize=.5\hsize}X}
    \hline\hline
        Inputs & $\braket{x}$ Frequency &  $\braket{x}$ RMSE &  $\braket{x^2}$ RMSE \\ \hline 
        $\braket{x}$, $\braket{x^2}$, $(\frac{\partial \braket{x}}{\partial t})^2$, and $(\frac{\partial S}{\partial t})^2$ & $1.40243$ & $3.41483\times 10^{-4}$ & $1.87218\times 10^{-4}$ \\ \
        $\braket{x}$, $\braket{x^2}$, and $(\frac{\partial S}{\partial t})^2$ & $1.40260$  & $2.86096 \times 10^{-4}$ & $2.67096  \times 10^{-4}$  \\ 
        $\braket{x}$, $\braket{x^2}$, and $(\frac{\partial \braket{x}}{\partial t})^2$ & $1.40266$ & $1.54922 \times 10^{-4}$ & $3.39435 \times 10^{-4}$  \\ 
        $\braket{x}$ and $\braket{x^2}$ & $1.40249$ & $2.36722  \times 10^{-4}$ & $3.67558 \times 10^{-4}$ \\ \hline\hline
    \end{tabularx}
    \caption{
    Root-mean-square error (RMSE) for the first-order and the second-order moments, together with the oscillation frequency of the first-order moment in the MPT simulations with different input descriptors in the ANN model.
    }
    \label{table:RMSE}
\end{table}

\section{Conclusion}
We presented a novel theoretical formulation of the quantum dynamics in terms of the moments within the single-particle description. 
The aim of propagating the moments instead of single-particle wave functions is to reduce the high computational cost by expressing the quantum dynamics in terms of increasing orders of the moments. By doing so, the equation of motion for the quantum dynamics resembles that of classical molecular dynamics, and therefore modern machine-learning techniques might be employed more effectively.
We derived analytical expressions for the first-order and second-order time derivatives of the moments, and they are used to develop a numerical method for performing the quantum dynamics by expanding the moments in the Taylor series as done similarly for atom positions in classical molecular dynamics simulation. 
We demonstrated a few numerical schemes using the newly-developed moment propagation theory (MPT) on the harmonic oscillator potential, for which an exact analytical solution can be derived. We also applied the MPT method to an anharmonic Morse potential and compared it to the standard approach of propagating the wave function according to the time-dependent Schrodinger equation. 

The new moment propagation theory (MPT) offers a particularly compact representation suitable for developing machine-learned models for performing efficient quantum dynamics simulations.  
We demonstrated the use of a simple artificial neural network (ANN) model for machine-learning the second-order time derivatives of the moments so that their explicit evaluation can be circumvented in numerical simulations. 
Quantum dynamics simulation of electrons in molecules and condensed matter systems 
is where this new approach can find its immediate application. 
The new MPT framework is particularly powerful in practice when only low orders of the moments are necessary. 
In the context of electronic structure theory, real-time time-dependent density functional theory (RT-TDDFT) simulation can be cast in the Wannier gauge for simulating electron dynamics\cite{doi:10.1063/5.0057587,10.1063/1.5095631}. The maximally-localized Wannier functions (MLWFs) are spatially localized single-particle orbitals, and the MPT is likely to offer a convenient theoretical framework for developing machine-learning models for RT-TDDFT simulations performed in the Wannier gauge. 
As for the scaling of the MPT method with respect to the order of moments, the computational cost per each MLWF scales as $\frac{1}{6}n^3 + n^2+ \frac{11}{6}n$, where $n$ is the highest order of moments included. The computational scales linearly with the number of MLWFs in a system.
Having the formal theoretical formulation established in this work, future work will explore the use of this new MPT in the context of first-principles electronic structure theory for studying real chemical systems. 
In particular, the use of ANNs and other related machine-learning methods for simulating electron dynamics will be examined in the context of RT-TDDFT simulation.

\begin{acknowledgments}
N.B. was supported by the Chapel Family summer research fellowship. This work was supported by the National Science Foundation, under No. CHE-1954894.
\end{acknowledgments}

\section*{Author Declarations}

\subsection*{Conflict of Interest}

The authors have no conflicts to disclose.

\subsection*{Author Contributions}
N.B. led the work and performed all the calculations. N.B. and Y.K. conceived of the presented idea. All authors discussed the results and contributed to the final manuscript.

\subsection*{Data Availability}
The data that support the findings of this study are available from the corresponding author upon reasonable request.

\appendix
\setcounter{equation}{0}
\renewcommand{\theequation}{A\arabic{equation}}
\setcounter{subsection}{0}
\renewcommand{\thesubsection}{A\arabic{subsection}}

\section{Notations}

We used the following notations to define several gradient operations in Appendix,

\begin{equation}
\nabla = (\partial_x, \partial_y, \partial_z),
\end{equation}

where $\partial_x \equiv \frac{\partial}{\partial x}$.
\noindent
Accordingly,
\begin{equation}
\begin{aligned}
\nabla^2 &= \nabla \cdot \nabla, \\  
\nabla^3 &= \nabla \nabla \cdot \nabla, \\  
\nabla^4 &= (\nabla \cdot\nabla )(\nabla \cdot \nabla ).\\ 
\end{aligned}
\end{equation}
$\nabla^2 $ is the Laplacian operator, and therefore we have
\begin{equation}
\nabla^2 (  \nabla A  \cdot \nabla B) = \nabla^3 A  \cdot \nabla B +2 (\nabla \otimes \nabla A ) \cdot (\nabla \otimes \nabla B)+\nabla A  \cdot \nabla^3 B,
\end{equation}
where tensor product of $\nabla$ is
\begin{equation}
\nabla \otimes \nabla = (\partial_{xx}, \partial_{xy}, \partial_{xz}, \partial_{xy}, \partial_{yy}, \partial_{zy}, \partial_{xz}, \partial_{zy}, \partial_{zz}) .
\end{equation}

\noindent

\section{Derivation of Eq. \ref{eq:firstder}}
\setcounter{equation}{0}
\renewcommand{\theequation}{B\arabic{equation}}
\setcounter{subsection}{0}
\renewcommand{\thesubsection}{B\arabic{subsection}}
\label{sec:eq3derivation}

For a single-particle system in three dimensions, the derivative with respect to time for the moment is defined as:
\begin{equation}
    \label{eq:ftd}
    \frac{d \braket{x^ay^bz^c}}{d t} \equiv \int x^ay^bz^c \frac{d n(x, y, z, t)} {d t} d^3r.
\end{equation}
Here we used $\int d^3r$ to indicate the spacial integral $\int\int\int dx dy dz$ 
for brevity.
Using $n(x,y,z,t)=\psi^*(x,y,z,t)\psi(x,y,z,t)=|\psi(x,y,z,t)|^2$ and the TD-SE (Eq. \ref{eq:tdse}), one has
\begin{equation}
    \begin{aligned}
        \frac{d \braket{x^ay^bz^c}}{dt}
        = & \int x^ay^bz^c \left[\left(\frac{i}{2}\nabla^2\psi - iV\psi\right)\psi^* + \psi\left(-\frac{i}{2}\nabla^2\psi^* + iV\psi^*\right)\right]d^3r \\
        = & \frac{i}{2}\int x^ay^bz^c \left(\nabla^2\psi\psi^* - \psi\nabla^2\psi^*\right) d^3r \\
        = & -\frac{i}{2}\int \nabla\left(x^ay^bz^c\right)\cdot\left(\nabla\psi\psi^* - \psi\nabla\psi^*\right) d^3r \\
        = & -\frac{i}{2} \int \left[\nabla^2\left(x^ay^bz^c\right)n + 2\nabla\left(x^ay^bz^c\right)\cdot\nabla\psi\psi^*\right]d^3r,\\
    \end{aligned}
    \label{eq:fdfull}
\end{equation}

where the explicit spatial and time dependence of the wave function $\psi(x,y,z,t)$, time-dependent potential $V(x,y,z,t)$, and density $n(x,y,z,t)$ are omitted for brevity. The explicit expression is:
\begin{equation}
    \begin{aligned}
        \frac{d \braket{x^ay^bz^c}}{d t} = & -i \int \int \int (a x^{a-1}y^bz^c \partial_x\psi\psi^*+ b y^{b-1}x^az^c \partial_y\psi\psi^* + c z^{c-1}x^ay^b \partial_z\psi\psi^* \\
        + &\frac{a(a-1)}{2} x^{a-2}y^bz^c n(x, y, z, t) + \frac{b(b-1)}{2} y^{b-2}x^az^c n(x, y, z, t) \\ 
        + &\frac{c(c-1)}{2} z^{c-2}x^ay^b n(x, y, z, t)) dx dy dz.
    \end{aligned}
\end{equation}
Using $\nabla\psi\psi^*-\psi\nabla\psi^* = 2i \operatorname{Im}(\nabla\psi\psi^*)$, Eq. \ref{eq:fdfull} can be also written concisely as
\begin{equation}
    \frac{d \braket{x^ay^bz^c}}{d t} = \operatorname{Re}\left[-i\int \nabla\left(x^ay^bz^c\right)\cdot\nabla\psi\psi^* d^3r\right].
    \label{eq:fd}
\end{equation}

\section{Derivation of Eq. \ref{eq:secondder}}
\setcounter{equation}{0}
\renewcommand{\theequation}{C\arabic{equation}}
\setcounter{subsection}{0}
\renewcommand{\thesubsection}{C\arabic{subsection}}
\label{sec:eq4derivation}

\noindent
Given Eq. \ref{eq:fd}, the second-order time derivative of the moment is:
\begin{equation}
\label{eq:a31}
    \frac{d^2 \braket{x^ay^bz^c}}{d t^2} = \operatorname{Re}\left[-i\frac{d}{d t}\int \nabla\left(x^ay^bz^c\right)\cdot\nabla\psi\psi^* d^3r\right].
\end{equation}

For the rest of the derivations in this section, the real part is implied. 
Analogously to the previous section, $n(x,y,z,t)=|\psi(x,y,z,t)|^2$ and TD-SE (Eq.\ref{eq:tdse}) gives,
\begin{equation}
    \begin{aligned}
        \frac{d^2 \braket{x^ay^bz^c}}{d t^2} 
        = & -i\int \nabla\left(x^ay^bz^c\right)\cdot\left(\nabla\frac{\partial \psi}{\partial t}\psi^* + \nabla\psi\frac{\partial \psi^*}{\partial t} \right)d^3r \\
        = & -i\int \nabla\left(x^ay^bz^c\right)\cdot\left[\nabla\left(\frac{i}{2}\nabla^2\psi - iV\psi\right)\psi^* + \nabla\psi\left(-\frac{i}{2}\nabla^2\psi^* + iV\psi^*\right)\right]d^3r \\
        = & \int \nabla\left(x^ay^bz^c\right)\cdot\left[-\left(\nabla V\right)n + \frac{1}{2}\left(\nabla^3\psi\psi^* - \nabla\psi\nabla^2\psi^*\right)\right]d^3r. \\
    \end{aligned}
    \label{eq:sdlong}
\end{equation}

\noindent
Expanding Eq. \ref{eq:sdlong}, we have
\begin{equation}
\frac{d^2 \braket{x^ay^bz^c}}{d t^2} = 
-\int \nabla (x^ay^bz^c) \cdot \nabla V n d^3 r+ \int \nabla\left(x^ay^bz^c\right)\cdot\frac{1}{2}(\nabla^3\psi\psi^* - \nabla\psi\nabla^2\psi^*)d^3r.
\end{equation}

\noindent
We now expand the right-hand side and apply the integration by parts,

\begin{equation}
\label{eq:Edef1}
\begin{aligned}
\frac{d^2 <x^ay^bz^c>}{d t^2} =
& 
-\int \nabla (x^ay^bz^c) \cdot \nabla V n d^3 r+ \int \nabla\left(x^ay^bz^c\right)\cdot\frac{1}{2}(\nabla^3\psi\psi^*)d^3r \\
&-\frac{1}{2}\int \nabla^2 (  \nabla \left(x^ay^bz^c\right)  \cdot \nabla\psi)  \psi^* d^3r\\
=& 
-\int \nabla (x^ay^bz^c) \cdot \nabla V n d^3 r-\frac{1}{2}\int  (  \nabla^3 \left(x^ay^bz^c\right)  \cdot \nabla\psi \\
&+2 \nabla \otimes \nabla \left(x^ay^bz^c\right)  \cdot (\nabla \otimes \nabla \psi))  \psi^* d^3r .\\
\end{aligned}
\end{equation}
\noindent
To make further progress on this equation to the working form, we used the real part of Eq. \ref{eq:fdfull},

\begin{equation}
\operatorname{Re}\left(\int\nabla\left(x^ay^bz^c\right)\cdot\nabla\psi\psi^*d^3r\right) = -\frac{1}{2} \int \nabla^2\left(x^ay^bz^c\right)n d^3r.
\end{equation}
\noindent
Then, Eq. \ref{eq:Edef1} is expressed as
\begin{equation}
\frac{d^2 \braket{x^ay^bz^c}}{d t^2} = \int \operatorname{Re} \left[ - \nabla (x^ay^bz^c) \cdot \nabla V n  + \frac{1}{4} \nabla^4\left(x^ay^bz^c\right)n  -  ( \nabla \otimes \nabla \left(x^ay^bz^c\right)  \cdot (\nabla \otimes \nabla \psi))  \psi^* \right] d^3r,
\end{equation}
\noindent
or completely expanded as
\begin{equation}
\begin{aligned}
\frac{d^2 \braket{x^ay^bz^c}}{d t^2} = & \operatorname{Re}(
- \int \int \int (
ax^{a-1}y^bz^c\partial_x V(x, y, z, t) n(x, y, z, t) + b x^ay^{b-1}z^c\partial_y V(x, y, z, t) n(x, y, z, t)\\
+ & cx^ay^bz^{c-1}\partial_z V(x, y, z, t) n(x, y, z, t))d^3r\\
- & \int \int \int 
(
a(a-1) x^{a-2}y^bz^c   \partial_x^2  \psi  \psi^* 
 + 2ba x^{a-1}y^{b-1}z^c   \partial_x \partial_y \psi  \psi^* \\
+ & 2ca x^{a-1}y^bz^{c-1}   \partial_x \partial_z \psi  \psi^* 
+ b(b-1) y^{b-2}x^az^c \partial_y^2  \psi  \psi^*\\ 
+ & 2cb y^{b-1}x^az^{c-1}  \partial_y \partial_z \psi  \psi^* 
+ c(c-1) z^{c-2}x^ay^b \partial_z^2  \psi \psi^*
)d^3r\\
+ &\frac{1}{4} \left( a(a-1) (
(a-2)(a-3) \braket{x^{a-4}y^bz^c} 
+ 2b(b-1) \braket{y^{b-2}x^{a-2}z^c} \right.\\
+ & 2c(c-1) \braket{z^{c-2}x^{a-2}y^b} 
) \\
+ & b(b-1) ( (b-2)(b-3) \braket{y^{b-4}x^az^c}
+ 2c(c-1) \braket{z^{c-2}x^ay^{b-2}} 
) \\
+ & \left.c(c-1)(c-2)(c-3) \braket{z^{c-4}x^ay^b}
 \right)) .
 \\
\end{aligned}
\label{eq:fullsda}
\end{equation}

\section{Derivation of Eq. \ref{eq:wfphase}}
\setcounter{equation}{0}
\renewcommand{\theequation}{D\arabic{equation}}
\setcounter{subsection}{0}
\renewcommand{\thesubsection}{D\arabic{subsection}}
\label{sec:eq13derivation}

The phase of the wave function in the polar form can be expressed as 
$\theta = atan2 (\operatorname{Im} \psi, \operatorname{Re} \psi)$, and the real and imaginary parts of the wave function are $\operatorname{Re} \psi = \sqrt{n} \cos{\theta}$ and $\operatorname{Im} \psi = \sqrt{n} \sin{\theta}$. 
Using the definition (from the main text) $\alpha_u \equiv \frac{i}{2} (  \partial_u^2\psi\psi^* - \psi\partial_u^2\psi^* )$, let us write 
\begin{equation}
L_u(x,y,z,t) = -\int_{- \infty}^u \alpha_{u}(\mathbf{r}', t) du' = \frac{-i}{2} ( \frac{\partial \psi}{ \partial u} \psi^* -\frac{\partial \psi^*}{ \partial u} \psi)
\end{equation}

where $u$ can be $x$, $y$, or $z$ while $\textbf{r}'$ denotes the dependence of the function on all three coordinates for brevity. 
We can rearrange this expression by writing out the real and imaginary parts of the right-hand side
 \begin{equation}
 \begin{aligned}
    L_u &= \frac{-i}{2} ( \frac{\partial \psi}{ \partial u} \psi^* -\frac{\partial \psi^*}{ \partial u} \psi) \\
    &= \frac{-i}{2} (i \operatorname{Im}(\frac{\partial \psi}{\partial u} \psi^*) +  \operatorname{Re}(\frac{\partial \psi}{\partial u} \psi^*) -i \operatorname{Im}(\frac{\partial \psi^*}{ \partial u} \psi)) -  \operatorname{Re}(\frac{\partial \psi^*}{ \partial u} \psi)))\\
    &=\frac{-i}{2} (2i \operatorname{Im}(\frac{\partial \psi}{\partial u} \psi^*))\\
    &= \frac{\partial \operatorname{Im} \psi}{\partial u } \operatorname{Re} \psi - \frac{\partial \operatorname{Re} \psi}{\partial u } \operatorname{Im} \psi .\\
\end{aligned}
 \label{eq:Lr}
 \end{equation}

The derivative of the phase can be expressed as 
\begin{equation}
\label{eq_dtdr}
\frac{\partial \theta}{\partial u } = -\frac{\operatorname{Im} \psi \frac{\partial \operatorname{Re} \psi}{\partial u}}{(\operatorname{Im} \psi)^2 + (\operatorname{Re} \psi)^2} + \frac{\operatorname{Re} \psi \frac{\partial \operatorname{Im} \psi}{\partial u}}{(\operatorname{Im} \psi)^2 + (\operatorname{Re} \psi)^2} .
\end{equation}
Using $n = (\operatorname{Im} \psi)^2 + (\operatorname{Re} \psi)^2$ and Eq. \ref{eq:Lr}, we have
$\frac{\partial \theta}{\partial u } = \frac{L_u}{n}$
and thus 
\begin{equation}
\label{dLdn}
\frac{\partial \theta(x, y,z ,t)}{\partial u } = \frac{L_u(x, y,z ,t)}{n(x, y,z ,t)}= - (n(x, y,z ,t))^{-1}\int_{- \infty}^u \alpha_{u}(\mathbf{r}', t) du', \\
\end{equation}

where $\alpha_u \equiv \frac{i}{2} (\frac{\partial^2 \psi}{\partial u^2}\psi^* - \frac{\partial^2 \psi^*}{\partial u^2}\psi) $. 
For each coordinate variable $u$, we have

\begin{equation}
\begin{aligned}
\frac{\partial \theta(x, y,z ,t)}{\partial x } = \frac{L_x(x, y,z ,t)}{n(x, y,z ,t)}= - (n(x, y,z ,t))^{-1}\int_{- \infty}^x \alpha_{x}(x', y, z, t) dx' \\
\frac{\partial \theta(x, y,z ,t)}{\partial y } = \frac{L_y(x, y,z ,t)}{n(x, y,z ,t)}= - (n(x, y,z ,t))^{-1}\int_{- \infty}^y \alpha_{y}(x, y', z, t) dy' \\
\frac{\partial \theta(x, y,z ,t)}{\partial z } = \frac{L_z(x, y,z ,t)}{n(x, y,z ,t)}= - (n(x, y,z ,t))^{-1}\int_{- \infty}^z \alpha_{z}(x, y, z', t) dz' .\\
\end{aligned}
\end{equation}

\section{Derivation of Eq. \ref{eq:alphax}}
\setcounter{equation}{0}
\renewcommand{\theequation}{E\arabic{equation}}
\setcounter{subsection}{0}
\renewcommand{\thesubsection}{E\arabic{subsection}}
\label{sec:eq15derivation}
\par 
We start by expressing the wave function in polar form as $\psi(x, y, z, t)= \sqrt{n(x, y, z, t)} e ^{i \theta(x, y, z, t)}$. While $n(x,y,z,t)$ can be obtained from the Edgeworth series of the moments straightforwardly, the expression for the phase in terms of the moments is required.
Using the TD-SE, the first-order time-derivative of the particle density $n$ can be expressed as

\begin{equation}
\begin{aligned}
\label{ntalpha}
\frac{d n}{d t} &= (\frac{i}{2}\nabla^2\psi - iV\psi)\psi^* + \psi(-\frac{i}{2}\nabla^2\psi^* + iV\psi^*)\\
&= \frac{i}{2} (\partial_x^2 \psi \psi^* - \psi\partial_x^2\psi^* + \partial_y^2\psi\psi^*  -  \psi\partial_y^2\psi^* + \partial_z^2 \psi \psi^* -  \psi\partial_z^2 \psi^*)\\
& = \alpha_x + \alpha_y + \alpha_z,
\end{aligned}
\end{equation}
where $\alpha_u \equiv \frac{i}{2} (  \partial_u^2\psi\psi^* - \psi\partial_u^2\psi^* )$.
Inserting this expression into Eq. \ref{eq:ftd}, we obtain 

\begin{equation}
\label{mtalpha}
\frac{d \braket{x^ay^bz^c}}{d t} = \int  x^ay^bz^c
( \alpha_x + \alpha_y + \alpha_z) d^3r.\\
\end{equation}
By defining 
\begin{equation}
\label{eq:cdef}
c_u^{abc}(t) \equiv \left(\frac{d \braket{x^ay^bz^c}}{d t}\right)^{-1} \int x^ay^bz^c
( \alpha_u(x, y, z,t)) d^3r,\\
\end{equation}

and Eq. \ref{mtalpha} can be expressed as 
\begin{equation}
\label{mtcmt}
\frac{d \braket{x^ay^bz^c}}{d t} = c_x^{abc}(t) \frac{d \braket{x^ay^bz^c}}{d t} + c_y^{abc}(t) \frac{d \braket{x^ay^bz^c}}{d t}+ c_z^{abc}(t) \frac{d \braket{x^ay^bz^c}}{d t}.
\end{equation}
\noindent
The time-derivative of the particle density can be also rearranged to read
\begin{equation}
\label{dn_dt}
\frac{d n }{d t} = \sum_{a,b,c} \frac{\partial n}{\partial \braket{x^ay^bz^c}} \frac{d \braket{x^ay^bz^c}}{d t}.
\end{equation}
By substituting the left-hand side of this equation with Eq. \ref{ntalpha} and the time-derivative on the right-hand side  with Eq. \ref{mtcmt}, Eq. \ref{dn_dt} reads

\begin{equation}
\alpha_x + \alpha_y + \alpha_z = \sum_{a,b,c} \frac{\partial n}{\partial \braket{x^ay^bz^c}} 
\{
 c_x^{abc}(t) \frac{d \braket{x^ay^bz^c}}{d t} + c_y^{abc}(t) \frac{d \braket{x^ay^bz^c}}{d t}+ c_z^{abc}(t) \frac{d \braket{x^ay^bz^c}}{d t}
 \}
 .
\end{equation}
Thus, for the spatial derivative in 
each Cartesian direction, we have Eq. \ref{eq:alphax}
\begin{equation}
\alpha_u = \sum_{a,b,c} c_u^{abc}(t)
\frac{\partial n}{\partial \braket{x^ay^bz^c}} \frac{d \braket{x^ay^bz^c}}{d t}.
\label{eq:alphaxdef}
\end{equation}

\section{Derivation of Eq. \ref{eq:secondderMPT2up}}
\setcounter{equation}{0}
\renewcommand{\theequation}{F\arabic{equation}}
\setcounter{subsection}{0}
\renewcommand{\thesubsection}{F\arabic{subsection}}
\label{subsec:eq19}

For brevity, let us derive the equation for the one-dimensional case (using $x$) in detail.
The first step in the derivation is to show
\begin{equation}
\label{eq:wfreplace}
Re(\psi^*(x, t)\frac{\partial^2 \psi(x, t)}{\partial x^2} ) = (\frac{1}{2}\frac{\partial^2 n(x, t)}{\partial x^2} - \frac{(\frac{ \partial n(x, t)}{\partial x})^2}{4n(x, t)} - \frac{L(x, t)^2}{n(x, t)}) ,
\end{equation}

where the right-hand side is expressed in terms of the particle density and the moments without explicit dependence on the wave function. Let us first write the left-hand side of Eq. \ref{eq:wfreplace} as
\begin{equation}
\begin{aligned}
LHS =& \operatorname{Re}(\operatorname{Re} \psi(x, t)\frac{\partial^2 \operatorname{Re} \psi(x, t)}{\partial x^2} + \operatorname{Im} \psi(x, t)\frac{\partial^2 \operatorname{Im} \psi(x, t)}{\partial x^2} \\&+ i\operatorname{Re} \psi(x, t)\frac{\partial^2 \operatorname{Im} \psi(x, t)}{\partial x^2} -i\operatorname{Im} \psi(x, t)\frac{\partial^2 \operatorname{Re} \psi(x, t)}{\partial x^2}) \\
=& \operatorname{Re} \psi(x, t)\frac{\partial^2 \operatorname{Re} \psi(x, t)}{\partial x^2} + \operatorname{Im} \psi(x, t)\frac{\partial^2 \operatorname{Im} \psi(x, t)}{\partial x^2} .
\end{aligned}
\label{eq:a51}
\end{equation}

\noindent
To prove Eq. \ref{eq:wfreplace}, we  also relate the first and second derivatives of $n(x,t)$ to the wave function,

\begin{equation}
\label{eq:dndx}
\frac{\partial n(x)}{\partial x} = 2 \operatorname{Re} \psi(x) \frac{\partial \operatorname{Re} \psi(x)}{\partial x} + 2 \operatorname{Im} \psi(x) \frac{\partial \operatorname{Im} \psi(x)}{\partial x} ,
\end{equation}

\begin{equation}
\label{eq:d2ndx2}
\frac{\partial^2 n(x)}{\partial x^2} = 2 \operatorname{Re} \psi(x) \frac{\partial^2 \operatorname{Re} \psi(x)}{\partial x^2} +2  (\frac{\partial \operatorname{Re} \psi(x)}{\partial x})^2 + 2 \operatorname{Im} \psi(x) \frac{\partial^2 \operatorname{Im} \psi(x)}{\partial x^2} +2  (\frac{\partial \operatorname{Im} \psi(x)}{\partial x})^2.\\
\end{equation}

\noindent
Using Eq. \ref{eq_dtdr} as well as the above Eqs.\ref{eq:dndx}/\ref{eq:d2ndx2}, the right-hand side of Eq. \ref{eq:wfreplace} 
is 
\begin{equation}
\begin{aligned}
RHS
&=  \operatorname{Re} \psi(x) \frac{\partial^2 \operatorname{Re} \psi(x)}{\partial x^2} +  (\frac{\partial \operatorname{Re} \psi(x)}{\partial x})^2 
+\operatorname{Im} \psi(x) \frac{\partial^2 \operatorname{Im} \psi(x)}{\partial x^2} +  (\frac{\partial \operatorname{Im} \psi(x)}{\partial x})^2) \\
&- \frac{(\operatorname{Re} \psi(x) \frac{\partial \operatorname{Re} \psi(x)}{\partial x})^2 +2(\operatorname{Re} \psi(x) \frac{\partial \operatorname{Re} \psi(x)}{\partial x})(\operatorname{Im} \psi(x) \frac{\partial \operatorname{Im} \psi(x)}{\partial x}) +(\operatorname{Im} \psi(x) \frac{\partial \operatorname{Im} \psi(x)}{\partial x})^2 }{n(x)} \\
&- \frac{(\operatorname{Re} \psi(x) \frac{\partial \operatorname{Im} \psi(x)}{\partial x})^2 -2(\operatorname{Re} \psi(x) \frac{\partial \operatorname{Re} \psi(x)}{\partial x})(\operatorname{Im} \psi(x) \frac{\partial \operatorname{Im} \psi(x)}{\partial x}) +(\operatorname{Im} \psi(x) \frac{\partial \operatorname{Re} \psi(x)}{\partial x})^2 }{n(x)} \\
&=  \operatorname{Re} \psi(x) \frac{\partial^2 \operatorname{Re} \psi(x)}{\partial x^2} +  (\frac{\partial \operatorname{Re} \psi(x)}{\partial x})^2 
+\operatorname{Im} \psi(x) \frac{\partial^2 \operatorname{Im} \psi(x)}{\partial x^2} +  (\frac{\partial \operatorname{Im} \psi(x)}{\partial x})^2) \\
&- \frac{(\operatorname{Re} \psi(x) \frac{\partial \operatorname{Re} \psi(x)}{\partial x})^2  +(\operatorname{Im} \psi(x) \frac{\partial \operatorname{Im} \psi(x)}{\partial x})^2 +(\operatorname{Re} \psi(x) \frac{\partial \operatorname{Im} \psi(x)}{\partial x})^2 +(\operatorname{Im} \psi(x) \frac{\partial \operatorname{Re} \psi(x)}{\partial x})^2 }{n(x)} \\
&=  \operatorname{Re} \psi(x) \frac{\partial^2 \operatorname{Re} \psi(x)}{\partial x^2} +  (\frac{\partial \operatorname{Re} \psi(x)}{\partial x})^2 
+\operatorname{Im} \psi(x) \frac{\partial^2 \operatorname{Im} \psi(x)}{\partial x^2} +  (\frac{\partial \operatorname{Im} \psi(x)}{\partial x})^2) \\
&- \frac{n(x) ((\frac{\partial \operatorname{Re} \psi(x)}{\partial x})^2 + (\frac{\partial \operatorname{Im} \psi(x)}{\partial x})^2 )}{n(x)} \\
&= \operatorname{Re} \psi(x) \frac{\partial^2 \operatorname{Re} \psi(x)}{\partial x^2}
+\operatorname{Im} \psi(x) \frac{\partial^2 \operatorname{Im} \psi(x)}{\partial x^2}.
\end{aligned}
\end{equation}
This is indeed the left-hand side of Eq. \ref{eq:wfreplace} (see Eq.\ref{eq:a51}), and thus Eq. \ref{eq:wfreplace} is proved.

\noindent
From Eq. \ref{eq:fullsda}, the second-order time-derivative of the moments  in a one-dimensional system can be written as 
\begin{equation}
\label{eq:1d2dmdt}
\begin{aligned}
\frac{d^2 \braket{x^a}}{d t^2} = & 
- \int \int \int (
ax^{a-1}\partial_x V(x, t) n(x, t))d^3r\\
- & \int \int \int 
a(a-1) x^{a-2}  \operatorname{Re}( \partial_x^2  \psi  \psi^* 
 )d^3r\\
+ &\frac{1}{4} a(a-1) (
(a-2)(a-3) \braket{x^{a-4}} .
 \\
\end{aligned}
\end{equation}
By substituting the integrand in the second term with Eq. \ref{eq:wfreplace}, Eq. \ref{eq:1d2dmdt} can be expressed without explicit reference to the wave function as 

\begin{equation}
\begin{aligned}
 \frac{d^2 \braket{x^a}}{d t^2}= - &a\int{
\frac{\partial  v(x)}{\partial  x} n(x, t) x^{a-1}d x} 
+a(a-1)\int{( \frac{(\frac{ \partial  n(x, t)}{\partial x})^2}{4n(x, t)} + \frac{L(x, t)^2}{n(x, t)}) x^{a-2}  d x}
\\ - & \frac{a(a-1)(a-2)(a-3)}{4} \braket{x^{a-4}}. 
\end{aligned}
\end{equation}

\noindent
Extending this expression to the general case (three-dimensional system) is tedious but straightforward, 

\begin{multline}
\label{eq:secondderMPT2}
\frac{d^2 \braket{x^ay^bz^c}(t)}{d t^2} =
- \int \int \int (ax^{a-1}y^bz^c\frac{\partial V(x, y, z, t)}{\partial x} n(x, y, z, t) \\
+ b x^ay^{b-1}z^c\frac{\partial V(x, y, z, t)}{\partial y} n(x, y, z, t)
+ cx^ay^bz^{c-1}\frac{\partial V(x, y, z, t)}{\partial z} n(x, y, z, t))d^3r\\ 
+\int \int \int (
a(a-1) x^{a-2}y^bz^c  (\frac{(\frac{\partial  n(x, y, z, t)}{\partial x}^2)}{4n(x, y, z, t)} + \frac{L_x(x, y,z,t)^2}{n(x, y, z, t)})
 \\+ 2ba x^{a-1}y^{b-1}z^c  (\frac{(\frac{\partial  n(x, y, z, t)}{\partial x}\frac{\partial  n(x, y, z, t)}{\partial y})}{4n(x, y, z, t)} + \frac{L_x(x, y,z,t)L_y(x, y,z,t)}{n(x, y, z, t)})\\
+ 2ca x^{a-1}y^bz^{c-1}  (\frac{(\frac{\partial  n(x, y, z, t)}{\partial x}\frac{\partial  n(x, y, z, t)}{\partial z})}{4n(x, y, z, t)} + \frac{L_x(x, y,z,t)L_z(x, y,z,t)}{n(x, y, z, t)})
\\+ b(b-1) y^{b-2}x^az^c(\frac{(\frac{\partial  n(x, y, z, t)}{\partial y}^2)}{4n(x, y, z, t)} + \frac{L_y(x, y,z,t)^2}{n(x, y, z, t)})
\\+ 2cb y^{b-1}x^az^{c-1} (\frac{(\frac{\partial  n(x, y, z, t)}{\partial z}\frac{\partial  n(x, y, z, t)}{\partial y})}{4n(x, y, z, t)} + \frac{L_z(x, y,z,t)L_y(x, y,z,t)}{n(x, y, z, t)})\\
+ c(c-1) z^{c-2}x^ay^b(\frac{(\frac{\partial  n(x, y, z, t)}{\partial z}^2)}{4n(x, y, z, t)} + \frac{L_z(x, y,z,t)^2}{n(x, y, z, t)})
)d^3r\\
- a(a-1) (
\frac{(a-2)(a-3)}{4} \braket{x^{a-4}y^bz^c} 
+ \frac{b(b-1)}{2} \braket{y^{b-2}x^{a-2}z^c} 
+ \frac{c(c-1)}{2} \braket{z^{c-2}x^{a-2}y^b} 
) \\
-b(b-1) (
\frac{(b-2)(b-3)}{4} \braket{y^{b-4}x^az^c}
+ \frac{c(c-1)}{2} \braket{z^{c-2}x^ay^{b-2}} 
) \\
-\frac{c(c-1)(c-2)(c-3)}{4} \braket{z^{c-4}x^ay^b},\\
\end{multline}
which is Eq. \ref{eq:secondderMPT2up}.

\section{Derivation of Eq. \ref{eq:secondderMPT2upcx}}
\setcounter{equation}{0}
\renewcommand{\theequation}{G\arabic{equation}}
\setcounter{subsection}{0}
\renewcommand{\thesubsection}{G\arabic{subsection}}
\label{sec:eq16derivation}

Eq. \ref{eq:secondderMPT2upcx} gives the equation of motion for the product term 
 $c_u^{abc}(t) \frac{\partial \braket{x^ay^bz^c}(t)}{\partial t}$
expressed in terms of the particle density and the moments without explicit dependence on the wave function. 
We start by writing Eq. \ref{eq:cdef} as
\begin{equation}
\begin{aligned}
c_u^{abc}(t)  \frac{d \braket{x^ay^bz^c}}{d t} &= \int \int \int x^ay^bz^c
( \alpha_u(x, y, z,t)) d^3r,\\ 
& = \int \int \int x^ay^bz^c
(\frac{i}{2} (  \partial_u^2\psi\psi^* - \psi\partial_u^2\psi^* )) d^3r\\
&=\operatorname{Re} \left[-i\int \int \int 
   \partial_u (x^ay^bz^c) \partial_u \psi \psi^* d^3r \right] .
\end{aligned}
\label{eq:cmoment}
\end{equation}
For the rest of the derivations here in this subsection, the real part is implied. 
Analogously to the previous sections, using $n(\mathbf{r},t)=|\psi(\mathbf{r},t)|^2$ and TD-SE (Eq.\ref{eq:tdse}) yields  the time-derivative as
\begin{equation}
    \begin{aligned}
        \frac{d }{d t}(c_u^{abc}(t) \frac{d \braket{x^ay^bz^c}(t)}{d t})
        = & -i\int \partial_u \left(x^ay^bz^c\right) \left(\partial_u \frac{\partial \psi}{\partial t}\psi^* + \partial_u \psi\frac{\partial \psi^*}{\partial t} \right)d^3r \\
        = & -i\int \partial_u \left(x^ay^bz^c\right)\left[\partial_u \left(\frac{i}{2}\nabla^2\psi - iV\psi\right)\psi^* + \partial_u \psi\left(-\frac{i}{2}\nabla^2\psi^* + iV\psi^*\right)\right]d^3r \\
        = & \int \partial_u \left(x^ay^bz^c\right)\left[-\left(\partial_u V\right)n + \frac{1}{2}\left(\nabla^2 \partial_u \psi\psi^* - \partial_u \psi\nabla^2\psi^*\right)\right]d^3r \\
        = & -\int \partial_u (x^ay^bz^c)  \partial_u V n d^3 r+ \int \partial_u \left(x^ay^bz^c\right)\frac{1}{2}(\nabla^2 \partial_u \psi\psi^* - \partial_u \psi\nabla^2\psi^*)d^3r.
    \end{aligned}
    \label{eq:sdlong2}
\end{equation}

\noindent
Applying the integration by parts to the right-hand side, we have
\begin{equation}
\label{eq:Edef}
\begin{aligned}
\frac{d }{d t}(c_u^{abc}(t) \frac{d \braket{x^ay^bz^c}(t)}{d t}) =
& 
-\int \partial_u (x^ay^bz^c)  \partial_u V n d^3 r+ \int \partial_u\left(x^ay^bz^c\right)\frac{1}{2}(\nabla^2 \partial_u \psi\psi^*)d^3r \\
&-\frac{1}{2}\int \nabla^2 (  \partial_u \left(x^ay^bz^c\right)  \partial_u \psi)  \psi^* d^3r\\
=& 
-\int \partial_u (x^ay^bz^c) \partial_u  V n d^3 r-\frac{1}{2}\int  (  \nabla^2 \partial_u \left(x^ay^bz^c\right)  \partial_u \psi \\
&+2 \nabla  \partial_u \left(x^ay^bz^c\right)  \cdot (\nabla \partial_u \psi))  \psi^* d^3r.
\end{aligned}
\end{equation}
\noindent

To make further progress on this equation above to the working form, let us express
write the following with the integration by parts
\begin{equation}
\begin{aligned}
\label{eq:dmoment}
\left(\int \nabla^2 \partial_u \left(x^ay^bz^c\right) \partial_u \psi\psi^*d^3r\right) &= - \int \nabla^2 \partial_u^2\left(x^ay^bz^c\right)n d^3r - \left(\int \nabla^2 \partial_u \left(x^ay^bz^c\right)  \psi \partial_u \psi^*d^3r\right)\\
\left(\int \nabla^2 \partial_u \left(x^ay^bz^c\right) (\partial_u \psi\psi^* + \psi \partial_u 
 \psi^*)d^3r\right) &= - \int 
\nabla^2\partial_u^2\left(x^ay^bz^c\right)n d^3r \\
 \left(\int \nabla^2\partial_u \left(x^ay^bz^c\right) 2\operatorname{Re} (\partial_u \psi\psi^* )d^3r\right) &= - \int \nabla^2\partial_u^2\left(x^ay^bz^c\right)n d^3r \\
 \operatorname{Re} \left(\int\nabla^2\partial_u \left(x^ay^bz^c\right)  (\partial_u \psi\psi^* )d^3r\right) &= - \frac{1}{2}\int \nabla^2\partial_u^2\left(x^ay^bz^c\right)n d^3r, \\
\end{aligned}
\end{equation}
\noindent
where we used $(\partial_u \psi\psi^* + \psi \partial_u \psi^*)=2\operatorname{Re} (\partial_u \psi\psi^* ) $.
Using the final expression Eq. \ref{eq:dmoment} for the second term of the RHS in Eq. \ref{eq:Edef}, Eq. \ref{eq:Edef} is expressed as

\begin{equation}
\begin{aligned}
\frac{d }{d t}(c_u^{abc}(t) \frac{d \braket{x^ay^bz^c}(t)}{d t}) =& \int \operatorname{Re} ( - \partial_u (x^ay^bz^c) \partial_u V n  + \frac{1}{4} \nabla^2 \partial_u^2(x^ay^bz^c)n \\ 
&-  ( \nabla  \partial_u (x^ay^bz^c)  \cdot (\nabla  \partial_u \psi))  \psi^* ) d^3r.
\end{aligned}
\end{equation}

\noindent
Using the same approach used to derive Eq. \ref{eq:wfreplace} in Sec.\ref{subsec:eq19}, Eq. \ref{eq:Edef} in terms of the density and moments is

\begin{multline}
\label{eq:secondderMPT2downcx}
\frac{d }{d t}(c_u^{abc}(t) \frac{d \braket{x^ay^bz^c}(t)}{d t}) =
- \int \int \int (\partial_u (x^ay^bz^c)\frac{\partial V(x, y, z, t)}{\partial u} n(x, y, z, t))d^3r \\
+\int \int \int (
\partial_u \partial_x (x^ay^bz^c)  (\frac{(\frac{\partial  n(x, y, z, t)}{\partial u} \frac{\partial  n(x, y, z, t)}{\partial x})}{4n(x, y, z, t)} + \frac{L_u(x, y,z,t) L_x(x, y,z,t)}{n(x, y, z, t)})
 \\+ \partial_u \partial_y (x^ay^bz^c)  (\frac{(\frac{\partial  n(x, y, z, t)}{\partial u}\frac{\partial  n(x, y, z, t)}{\partial y})}{4n(x, y, z, t)} + \frac{L_u(x, y,z,t)L_y(x, y,z,t)}{n(x, y, z, t)}) \\
 +\partial_u \partial_z (x^ay^bz^c)  (\frac{(\frac{\partial  n(x, y, z, t)}{\partial u}\frac{\partial  n(x, y, z, t)}{\partial z})}{4n(x, y, z, t)} + \frac{L_u(x, y,z,t)L_z(x, y,z,t)}{n(x, y, z, t)})) d^3r\\
- \frac{1}{4}\braket{ \nabla^2 \partial_u^2\left(x^ay^bz^c\right)}. \\
\end{multline}

\section{Derivation of Eq. \ref{eq:En}}
\setcounter{equation}{0}
\renewcommand{\theequation}{H\arabic{equation}}
\setcounter{subsection}{0}
\renewcommand{\thesubsection}{H\arabic{subsection}}
\label{sec:eq31derivation}
For the one-dimensional system, we have
\begin{equation}
\epsilon = \braket{\psi | \hat{H} | \psi} = \int  -\frac{1}{2}\psi^*(x)\frac{\partial^2 \psi(x)}{\partial x^2} + \psi^*(x) V(x) \psi(x) dx.
\end{equation}
Using Eq. \ref{eq:wfreplace}, the wave function dependence can be recast in terms of the probability density and the moments as

\begin{equation}
\label{eq:EnergyX}
\epsilon = \int V(x) n(x) dx +\frac{1}{2} \int   (\frac{(\frac{\partial n(x)}{\partial x}^2)}{4n(x)} + \frac{L(x)^2}{n(x)}) dx.
\end{equation}

\section{Analytical solution for simple harmonic potential (Eqs. \ref{eq:FMTD}, \ref{eq:SMTD}, and \ref{eq:EneTh})}
\setcounter{equation}{0}
\renewcommand{\theequation}{I\arabic{equation}}
\setcounter{subsection}{0}
\renewcommand{\thesubsection}{I\arabic{subsection}}
\label{sec:eq27derivation}
The harmonic potential case, $V(x, t)=x^2 + c(t)x$, can be solved analytically within the framework of the moment propagation theory (MPT). For this system, the second-order Edgeworth basis (E2) is sufficient\cite{Andrews_2016}, and 
it requires only the first-order and the second-order moments to be propagated.

\subsection{ Analytical Solution}
\label{subsub_as}

The MPT equation of motion for a 1D system is:

\begin{equation}
\begin{aligned}
\frac{d^2 \braket{x^a}}{d t^2}=& -a\int{
\frac{\partial  V(x)}{\partial  x} n(x, t) x^{a-1}d x} 
+a(a-1)\int{( \frac{(\frac{ \partial  n(x, t)}{\partial x})^2}{4n(x, t)} + \frac{L(x, t)^2}{n(x, t)}) x^{a-2}  d x}
\\ &-
\frac{a(a-1)(a-2)(a-3)}{4} \braket{x^{a-4}}.
\end{aligned}
\end{equation}

\noindent
For the first ($a$=1) and second ($a$=2) order moments, the second time derivative can be simplified as

\begin{equation}
\label{eq:sdfm}
\frac{d^2 \braket{x}}{d t^2}= -\int{
\frac{\partial  V(x)}{\partial  x} n(x, t)d x}, 
\end{equation}

\begin{equation}
\label{eq:sdsm}
\frac{d^2 \braket{x^2}}{d t^2}= -2\int{
\frac{\partial  V(x)}{\partial  x} n(x, t) x d x} 
+2\int{( \frac{(\frac{ \partial  n(x, t)}{\partial x})^2}{4n(x, t)} + \frac{L(x, t)^2}{n(x, t)})  d x}.
\end{equation}
With the second-order Edgeworth (E2) basis, the particle density has a Gaussian form  

\begin{equation}
n(x, t) = \frac{1}{\sqrt{2 \pi (\braket{x^2}(t)- (\braket{x}(t))^2)} } \exp{(-\frac{1}{2} \frac{(x - \braket{x}(t))^2}{\braket{x^2}(t)- (\braket{x}(t))^2})}.
\end{equation}
\noindent
The spatial derivative is
\begin{equation}
\frac{\partial  n(x, t)}{\partial x} = -n(x, t)( \frac{(x - \braket{x}(t))}{\braket{x^2}(t)- (\braket{x}(t))^2}),
\label{eq:dnterm}
\end{equation}
and the time derivative is
\begin{equation}
\begin{aligned}
\frac{\partial  n(x, t)}{\partial t} = &\frac{-\frac{\partial  \braket{x^2}(t)}{\partial t}+ 2 \braket{x}(t)\frac{\partial  \braket{x}(t)}{\partial t}}{2\sqrt{2 \pi (\braket{x^2}(t)- (\braket{x}(t))^2)^3} } \exp{(\frac{-1}{2} \frac{(x - \braket{x}(t))^2}{\braket{x^2}(t) - (\braket{x}(t))^2})} \\ 
&-\frac{1}{2} n(x, t) (\frac{2 (x- \braket{x}(t)) (-\frac{\partial  \braket{x}(t)}{\partial t})}{(\braket{x^2}(t)- (\braket{x}(t))^2)}
- \frac{(\frac{\partial  \braket{x^2}(t)}{\partial t}- 2 \braket{x}(t)\frac{\partial  \braket{x}(t)}{\partial t}) (x - \braket{x}(t))^2}{(\braket{x^2}(t)- (\braket{x}(t))^2)^2}
) .
\end{aligned}
\end{equation}
\noindent
Rewriting the time derivative as a polynomial, we have

\begin{equation}
\label{eq:dndt_poly}
\frac{\partial  n(x, t)}{\partial t} = k_1 n(x, t) + k_2 x n(x, t) + k_3 x^2 n(x, t),
\end{equation}
where $\{k_i\}$ are given by
\begin{multline}
k_1= \frac{-\frac{\partial  \braket{x^2}(t)}{\partial t}+ 2 \braket{x}(t)\frac{\partial  \braket{x}(t)}{\partial t}}{2 (\braket{x^2}(t)- (\braket{x}(t))^2) }
+ \frac{\braket{x}(t)(-\frac{\partial  \braket{x}(t)}{\partial t})}{(\braket{x^2}(t)- (\braket{x}(t))^2)}\\
+ \frac{1}{2} \frac{(\frac{\partial  \braket{x^2}(t)}{\partial t}- 2 \braket{x}(t)\frac{\partial  \braket{x}(t)}{\partial t}) (\braket{x}(t))^2}{(\braket{x^2}(t)- (\braket{x}(t))^2)^2},\\
k_2x = \frac{x(\frac{\partial  \braket{x}(t)}{\partial t})}{(\braket{x^2}(t)- (\braket{x}(t))^2)} + \frac{1}{2} \frac{(\frac{\partial  \braket{x^2}(t)}{\partial t}- 2 \braket{x}(t)\frac{\partial  \braket{x}(t)}{\partial t}) (-2x \braket{x}(t))}{(\braket{x^2}(t)- (\braket{x}(t))^2)^2},\\
k_3x^2 = \frac{1}{2} \frac{(\frac{\partial  \braket{x^2}(t)}{\partial t}- 2 \braket{x}(t)\frac{\partial  \braket{x}(t)}{\partial t}) (x^2)}{(\braket{x^2}(t)- (\braket{x}(t))^2)^2}.\\
\end{multline}

Let us define $S$ as the variance (spread) as 
\begin{equation}
S \equiv \braket{x^2} - \braket{x}^2.
\end{equation}
In order to simplify the time derivative using the variance, 
we use $\braket{x} = 0$ for this harmonic potential case without loss of generality due to the translational invariance.  Eq. \ref{eq:dndt_poly} is then 

\begin{equation}
\begin{aligned}
\frac{\partial  n(x, t)}{\partial t} = k_1 n(x, t) + k_2 x n(x, t) + k_3 x^2 n(x, t),\\
k_1 = \frac{-\dot S}{2S},
k_2 = \frac{\frac{\partial  \braket{x}(t)}{\partial t}}{S},
k_3 = \frac{\dot S}{ 2 S^2}.
\end{aligned}
\end{equation}

\noindent
Integrating the time derivative over the position x using the definition $L(x, t) \equiv -\int_{- \infty}^{x}\alpha_x(x',t) dx'$, we obtain 
\begin{equation}
L(x, t) = -(k_1 + S k_3) \Phi(x, t) +S k_2 n(x, t) +Sxk_3n(x, t),
\end{equation}
where $\Phi(x, t) \equiv \int_{- \infty}^{x} n(x', t) dx'$.
Because $
(k_1  + S k_3)= \frac{-\dot S}{2S} + \frac{\dot S}{ 2 S}=0$, we have
\begin{equation}
\begin{aligned}
L(x, t)& =  a_1 n(x, t) +a_2xn(x, t), \\
a_1& = \frac{\partial  \braket{x}(t)}{\partial t}, 
a_2 = \frac{\dot S}{2 S}.
\end{aligned}
\label{eq:Lterms}
\end{equation}
Substituting Eqs. \ref{eq:Lterms} and \ref{eq:dnterm} into the equation of motion (EOM) for the first-order and second-order moments (Eqs. \ref{eq:sdfm} and \ref{eq:sdsm}), we have
\begin{equation}
\begin{aligned}
\frac{d^2 \braket{x}}{d t^2} &= -\int{
\frac{\partial  (x^2+c(t)x)}{\partial  x} n(x, t)d x} = 
-2\braket{x}(t)-c(t),
\end{aligned}
\end{equation}

\begin{equation}
\begin{aligned}
\frac{d^2 \braket{x^2}}{d t^2}=& -2\int{
\frac{\partial  (x^2+c(t)x)}{\partial  x} n(x, t) x d x}  \\
&+ 2\int{( \frac{1}{4S^2}n(x, t)x^2 + a_1^2 n(x, t) + 2 a_1a_2xn(x, t) + a_2^2x^2n(x, t))  d x} \\
=& -2\int
(2x+c(t))n(x, t) x +2( (\frac{\partial  \braket{x}(t)}{\partial t})^2 n(x, t) +  2\frac{\partial  \braket{x}(t)}{\partial t} \frac{\dot S}{2 S} x n(x, t) \\ 
&+ ((\frac{\dot S}{2 S})^2 x^2 + \frac{1}{4S^2} )n(x, t))  d x. \\
&= -4\braket{x^2}(t)-2c(t)\braket{x}(t) 
+2( (\frac{\partial  \braket{x}(t)}{\partial t})^2 + \frac{\dot S^2}{4 S} + \frac{1}{4S} )) .
\end{aligned}
\end{equation}
\noindent
Succinctly put, the second-order time derivatives  of the moments read 
\begin{equation}
\frac{d^2 \braket{x}}{d t^2}= -2\braket{x}-c,
\end{equation}

\begin{equation}
\frac{d^2 \braket{x^2}}{d t^2}= -4\braket{x^2}-2c\braket{x} 
+2((\frac{(\frac{\partial  S}{\partial t})^2}{4S})
+ \frac{1}{4S} + (\frac{\partial  \braket{x}(t)}{\partial t} )^2 ).
\end{equation}
The single-particle energy  is $\epsilon=\braket{\hat{H}}$, and it is equivalent to the system's total energy for this one-dimensional harmonic oscillator model. The expression for the energy is obtained by using Eqs. \ref{eq:Lterms} and \ref{eq:dnterm} in Eq. \ref{eq:EnergyX}, resulting in

\begin{equation}
\label{eq:EnergyX2}
E= \epsilon = \braket{x^2} + c\braket{x} 
+\frac{1}{2}((\frac{(\frac{\partial  S}{\partial t})^2}{4S}) 
+ \frac{1}{4S} + (\frac{\partial  \braket{x}(t)}{\partial t} )^2 ).
\end{equation}

\subsection{ Analytical Solution via EPES approach}
EPES expression (Eq. \ref{eq:EPES}) holds if/when the Hamiltonian expectation value is a constant of motion in the quantum dynamics given by a suitable single-particle Hamiltonian.
Then, EPES approach is easier to work with by deriving the second-order time-derivatives of the cumulants rather than the moments directly. Note that the first cumulant is equivalent to the first-order moment. We first relate the second cumulant (which is the variance) and their time-derivatives to the moments by expressing them as

\begin{equation}
\begin{aligned}
\label{eq:s_x}
S &\equiv \braket{x^2} - \braket{x}^2,\\
\dot{S} &= \dot{\braket{x^2}} - 2 \braket{x}\dot{\braket{x}},\\
\Ddot{S} &= \Ddot{\braket{x^2}} - 2 \dot{\braket{x}}^2 - 2 \braket{x} \Ddot{\braket{x}}.\\
\end{aligned}
\end{equation}

\noindent
Using Eq. \ref{eq:s_x} and Eq. \ref{eq:EnergyX2}, the EPES equation of motion (Eq. \ref{eq:EPES}) for the first-order moment/cumulant can be written as

\begin{equation}
\begin{aligned}
\Ddot{\braket{x}} 
&= -\frac{\partial \epsilon(x_i, \dot{x}_i)}{ \partial  \braket{x}} 
\left(
\frac{\partial  \epsilon(x_i, \dot{x}_i)}{ \partial  \dot{\braket{x}}}
\right)^{-1}
\dot{\braket{x}} \\
 &= -2\braket{x}-c.
\end{aligned}
\end{equation}
Rewriting Eq. \ref{eq:EPES} in terms of the cumulant via Eq. \ref{eq:s_x}, 
the second-order time derivative of the second cumulant can be rearrange as

\begin{equation}
\Ddot{S} = -\frac{\partial  \epsilon(x_i, \dot{x_i})}{ \partial  S} 
\left(\frac{\partial  \epsilon(x_i, \dot{x_i})}{\partial \dot{S}}
\right)^{-1}
\dot{S}.
\end{equation}

\noindent
Then, substituting Eq. \ref{eq:EnergyX2} in this expression yields 

\begin{equation}
\begin{aligned}
\Ddot{S} &= -(1 - \frac{1}{2}  (\frac{1+\dot{S}^2}{4S^2})) 
\frac{4S}{\dot{S}}
\dot{S}
\\
 &= -4S + \frac{1}{2}  (\frac{1+\dot{S}^2}{S}).
\end{aligned}
\end{equation}

\noindent
Converting this expression given in cumulants back to the one in moments by using Eq. \ref{eq:s_x}, we have

\begin{equation}
\Ddot{\braket{x^2}} = -4 \braket{x^2}- 2c \braket{x} + 2  (\frac{1+\dot{S}^2}{4S} +\dot{\braket{x}}^2 ),\\
\end{equation}
which is the same as the one derived using the MPT without relying on the EPES expression (see Section \ref{subsub_as}).

\bibliography{MPT}

\end{document}